\documentclass[12pt,journal,draftclsnofoot,onecolumn]{IEEEtran}

\usepackage{cite}
\usepackage{url}
\usepackage{amssymb,amsmath}
\usepackage{array}
\usepackage{graphicx}
\usepackage{subfigure}
\usepackage{cite}
\usepackage{epsfig}
\usepackage{epic}
\usepackage{graphics}
\usepackage{amsmath}
\usepackage{multicol}
\usepackage{hyperref}
\usepackage{graphicx}
\usepackage{array}
\usepackage{multirow}
\usepackage{fancybox,dashbox}
\usepackage{cases}
\usepackage{caption2}

\begin{document}
%
\title{Duality of Channel Encoding and Decoding - Part I: Rate-1 Binary Convolutional Codes}

\author{Yonghui~Li,~\IEEEmembership{Senior~Member,~IEEE,}~Qimin~You,
        Soung~C.~Liew,~\IEEEmembership{Fellow,~IEEE,}~and
        Branka~Vucetic,~\IEEEmembership{Fellow,~IEEE}
\thanks{This work was presented in part at the IEEE International Symposium on Information Theory (ISIT) 2012, Boston, MA, July 2012.}
\thanks{Yonghui Li, Qimin You and Branka Vucetic are with School of Electrical and Information Engineering,
        University of Sydney, Sydney, NSW, 2006, Australia. Email: \{yonghui.li,~qimin.you,~branka.vucetic\}@sydney.edu.au}
\thanks{Soung C. Liew is with Department of Information Engineering, The Chinese University of Hong Kong, Hong Kong, Email:soung@ie.cuhk.edu.hk.}}

\maketitle

\begin{abstract}
In this paper, we revisit the forward, backward and bidirectional
Bahl-Cocke-Jelinek-Raviv (BCJR) soft-input soft-output (SISO)
maximum a posteriori probability (MAP) decoding process of rate-1
binary convolutional codes. From this we establish some interesting
explicit relationships between encoding and decoding of rate-1
convolutional codes. We observe that the forward and backward BCJR
SISO MAP decoders can be simply represented by their dual SISO
channel encoders using shift registers in the complex number field.
Similarly, the bidirectional MAP decoding can be implemented by
linearly combining the shift register contents of the dual SISO
encoders of the respective forward and backward decoders. The dual
encoder structures for various recursive and non-recursive rate-1
convolutional codes are derived.
\end{abstract}



\section{Introduction}
Convolutional codes were first introduced by Elias more than 50
years ago [1].  They have been widely used in various modern
communications systems, such as space and satellite communications,
cellular mobile, and digital video broadcasting. Its popularity
stems from its simple encoder structure, which can be implemented by
shift registers.

The main complexity associated with systems using convolutional
coding is situated in the decoder.  Decoding essentially consists of
finding an optimal path in a trellis based graph. Various decoding
algorithms have been developed to achieve the optimal decoding
performance in the most efficient manner. The Viterbi algorithm (VA)
has been known as a maximum-likehood (ML) decoding method, which
minimizes the sequence error rate [2-4]. It exhaustively searches
all states of the trellis over a fixed length window and finds a
most likely information sequence. In the standard VA, the decoder
produces hard-decision outputs, which are the estimates of
transmitted binary information symbols. In [5, 8], the VA is
modified to deliver not only the most-likely binary signal sequence,
but also the soft output containing the a posteriori probabilities
(APPs) of the transmitted binary symbols. The soft-output VA (SOVA)
is especially useful when decoding concatenated codes, such as turbo
codes, as it provides soft input for the next decoding stage and
thus improved performance.

There exists another class of non-linear decoding algorithms, called
maximum a posteriori probability (MAP) decoding. It was first
proposed by Bahl, Cocke, Jelinek and Raviv (BCJR) in 1974 [6]. It
performs symbol by symbol decoding and uses the symbol error rate as
the optimization criterion. Both the input and output of the decoder
are soft information signals. Compared to the VA, the
soft-input-soft-output (SISO) MAP can provide the optimal
symbol-by-symbol APP, and thus can fully exploit the full benefits
of soft-decision decoding in iterative decoding process of
concatenated codes.

The BCJR MAP decoding is a bi-directional decoding process,
consisting of a forward and a backward recursion process, which
dominates the main complexity of a decoder. In each direction, the
decoder infers the probabilities of current states and information
symbols based on the probabilities of the previous states in the
forward and backward trellis, the received signal, the channel state
and the a priori probabilities of the transmitted signals. The
complexity of forward and backward recursion exponentially increases
with the constraint length of convolutional codes.

In this paper, we revisit the forward, backward and bidirectional SISO MAP decoding of rate-1 convolutional codes. We observe some interesting explicit relationship between a SISO forward/backward MAP decoder of a convolutional code and its encoder. The forward and backward decoder of a rate-1 convolutional code can actually be represented by its corresponding dual encoder using shift registers in the complex field. This significantly reduces the original exponential computational complexity of MAP forward and backward recursion to the linear complexity. Similarly the bidirectional MAP decoding can be implemented by linearly combining
the shift register contents of the dual SISO encoders of the respective forward and backward decoders. With logarithm of the soft coded symbol estimate, directly obtained from the received signals, as the input to the dual encoder, the dual encoder output produces the logarithm of the soft symbol estimates of the binary information symbols.

We found that the dual encoder structure of a code depends on whether the code is recursive
or not. In our preliminary work in [9], we investigated the rate-1 recursive convolutional codes.
In this paper, we will study the general rate-1 convolutional codes, including the feedback only
convolutional (FBC) code, feed-forward only convolutional (FFC) code and general convolutional
(GC) code. We will investigate the explicit relationship between a SISO forward/backward MAP
decoder of these codes. The dual encoder structure is derived for each class of codes. In [9], the
bidirectional decoding output is derived through the linear combination of forward and backward
decoder outputs. These complex coefficients are found through computer search. However we
only found the coefficients for some specific 4-state and 8-states codes due to the high complexity
involved in the search. In this paper, we propose a simple and general combining approach to
represent the bidirectional MAP decoder by linearly combining shift register contents of the dual
encoders of the respective forward and backward decoders. We prove that such linear combining
produces exactly the same decoding output as the bidirectional MAP decoding for any rate-1
convolutional codes.

The remainder of the paper is organized as follows. In Section II,
we first briefly review the BCJR forward decoding algorithm and
derive the dual encoder structures of MAP forward decoders for three
classes of rate-1 convolutional codes. The dual encoder structure
for backward decoding is presented in Section III. The
representation of bidirectional MAP decoding by using the derived
dual encoder structures of forward and backward decoding is
described in Section IV. Simulation results are shown in Section V.
Conclusions are drawn in Section VI.

\section{Linear Representation of MAP Forward Decoding}

In this section, we first revisit the BCJR forward decoding
algorithm. We will focus on the decoding of a single constituent
convolutional code of rate-1. Let $\mathbf{b}=(b_1,b_2,\ldots,b_K)$
be a binary information symbol sequence to be transmitted, where $K$
is the frame length. Let $\mathbf{c}=(c_1,c_2,\ldots,c_K)$ be the
binary codeword of $\mathbf{b}$, generated by the binary code
generator polynomial $\mathbf{g}$, and
$\mathbf{x}=(x_1,x_2,\ldots,x_K)$ be the modulated symbol sequence
of $\mathbf{c}$. For simplicity, we consider the BPSK modulation.
Let $\mathbf{y}=(y_1,y_2,\ldots,y_K)$ denote the received signal
sequence at the channel output.

Based on the encoder structure, we define three different classes of
convolutional codes. Let $a(x)=x^n+a_{n-1}x^{n-1}+\cdots+a_1x_1+1$
and $q(x)=x^n+q_{n-1}x^{n-1}+\cdots+q_1x_1+1$, where $n$ is the
degree of polynomials $a(x)$ and $g(x)$. We define a convolutional
code, generated by $g_{FBC}(x)=1/q(x)$, as a feedback-only
convolutional (FBC) code, a code generated by $g_{FFC}(x)=a(x)$ as a
feed-forward only convolutional (FFC) code, and a code generated by
$g_{GC}(x)=a(x)/g(x)$, as a general convolutional (GC) code. We will
investigate the forward decoding process of these three classes of
convolutional codes.

\subsection{Forward decoding of a FBC code}

In this subsection, we first investigate the forward decoding of an
FBC code. To gain better insight into the decoding process, let us
first look at the following example.

\begin{figure}
\centering
\includegraphics[width=0.8\textwidth]{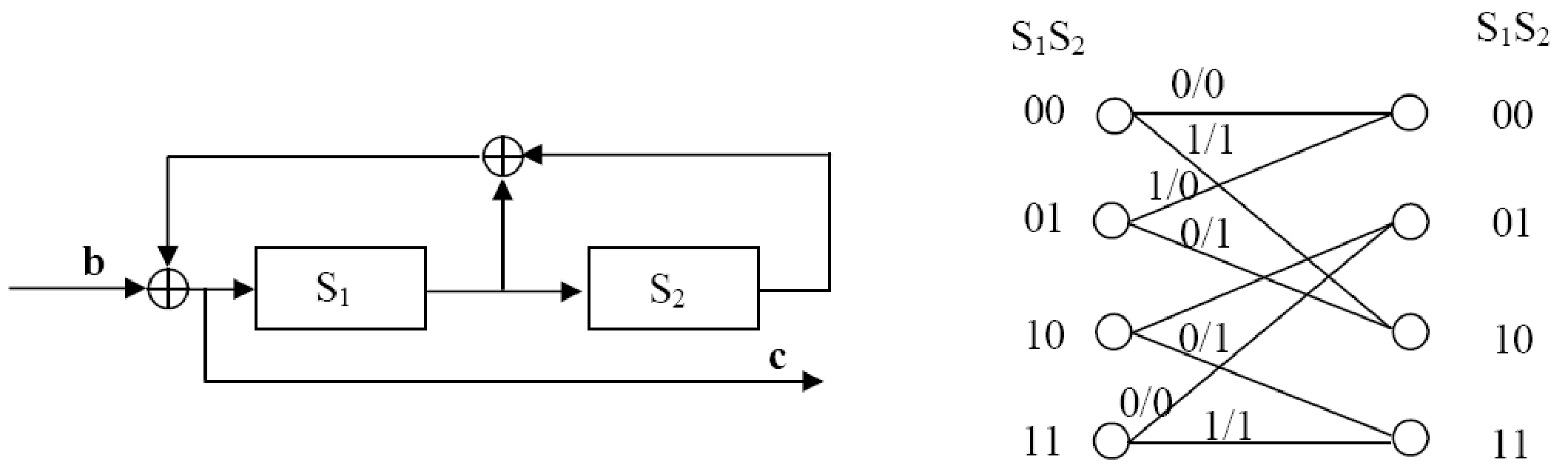}
\caption{The encoder and trellis of $g_{FBC}(x)=\frac{1}{x^2+x+1}$}
\label{fig0}
\end{figure}

\textbf{Example 1:} We consider a FBC code with the generator
polynomial of $g_{FBC}(x)=\frac{1}{x^2+x+1}$, for which the encoder
and trellis diagram are shown in Fig. \ref{fig0}. In the trellis
diagram, the state is labeled as $S_1S_2$ , where $S_i$, $i=1, 2$ is
the value of the $i$-th encoder shift register content. Each branch
in the trellis is labeled as $x/y$ where $x$ and $y$ denote the
encoder input and output, respectively.

Let  $p_{c_k}(l)=p(c_k=l|y_k)$, $l=0, 1$, denote the a posteriori
probabilities (APP) of the encoded symbol $c_k=l$, given the
received signal $y_k$, where $c_k$ is the transmitted binary coded
symbol at time $k$. Let us further denote
$\mathbf{P}_c=\{(p_{c_1}(0),p_{c_1}(1)),\cdots$,
$(p_{c_k}(0),p_{c_k}(1)),\cdots$, $(p_{c_K}(0),p_{c_K}(1))\}$. Now
let us follow the BCJR forward decoding algorithm to use
$\mathbf{P}_c$ to calculate the APPs of binary information symbols
$b_k$. Let $p_{b_k}(w)=p(b_k=w|\mathbf{y})$ represent the
probability of information symbol $b_k=w$, $w$=0, 1, given the
received signals $\mathbf{y}=\{y_1,\cdots,y_k,\cdots,y_K\}$. It can
be calculated in the following recursive way [6]
\begin{eqnarray} \label{eq1}
p_{b_k}(w)&=&p(b_k=w|\mathbf{y})=\sum_{(m',m)\in{U(b(k)=w)}}{\alpha_{k-1}(m')\gamma_k(m'm)}
 \\ \nonumber &=& \sum_{(m',m)\in{U(b(k)=w)}}{\alpha_{k-1}(m')p_{c_k}(c_k(m',m))}
\end{eqnarray}
\begin{equation} \label{eq2}
\alpha_k(m)=\sum_{m'}{\alpha_{k-1}(m')\gamma_k(m'm)}=\sum_{m'}{\alpha_{k-1}(m')p_{c_k}(c_k(m',m))},
\end{equation}
where $U(b(k)=w)$ is the set of trellis branches from the state $m'$
at time $k$-1 to the state $m$ at time $k$, that are caused by the
input binary symbol $b(k)=w$, and $c_k(m',m)$ represents the encoder
output of the corresponding trellis branch.

Let $m=0, 1, 2, 3$ represent the states of $S_1S_2=00, 01, 10, 11$
at time $k$, and
$\mathbf{\hat{x}_c}=(\hat{x}_{c_1},\cdots,\hat{x}_{c_K})$ and
$\mathbf{\hat{x}_b}=(\hat{x}_{b_1},\cdots,\hat{x}_{b_K})$ denote the
soft symbol estimate sequence of codeword $\mathbf{c}$ and
information sequence $\mathbf{b}$, respectively. We assume that 0
and 1 are modulated into symbol 1 and -1. Then the soft symbol
estimates $\hat{x}_{c_k}$ and $\hat{x}_{b_k}$, which represent the
probabilistic average of estimates of symbols ${x}_{c_k}$ and
${x}_{b_k}$ given $\mathbf{y}$, can be calculated as
\begin{eqnarray}
\hat{x}_{c_k}=E(x_{c_k}|y_k)=p_{c_k}(0)-p_{c_k}(1) \\
\hat{x}_{b_k}=E(x_{b_k}|\mathbf{y})=p_{b_k}(0)-p_{b_k}(1).
\end{eqnarray}

Then by using Eqs. (\ref{eq1}) and (\ref{eq2}) alternatively in
Example 1, we can get

(1) at time $k=0$,

$\alpha_0(0)=1$; $\alpha_0(1)=0$; $\alpha_0(2)=0$; $\alpha_0(3)=0$;

$p_{b_0}(0)=1$;~$p_{b_0}(1)=0$;

(2) at time $k=1$, the received signal is $y(1)$, and the input to
the decoder is the APPs of $c_1$, given by $p_{c_1}(0)$ and
$p_{c_1}(1)$, respectively. Then we have

$\alpha_1(0)=p_{c_1}(0)$; $\alpha_1(1)=0$; $\alpha_1(2)=p_{c_1}(1)$;
$\alpha_1(3)=0$;

$p_{b_1}(0)=p_{c_1}(0)$; $p_{b_1}(1)=p_{c_1}(1)$;

and

$\hat{x}_{b_1}=p_{b_1}(0)-p_{b_1}(1)=p_{c_1}(0)-p_{c_1}(1)=\dashbox{$\hat{x}_{c_1}$}$

(3) at time $k=2$, the input to the decoder is the APPs of $c_2$,
$p_{c_2}(0)$ and $p_{c_2}(1)$. We have

$\alpha_2(0)=p_{c_2}(0)p_{c_1}(0)$;
$\alpha_2(1)=p_{c_2}(0)p_{c_1}(1)$;
$\alpha_2(2)=p_{c_2}(1)p_{c_1}(0)$;
$\alpha_2(3)=p_{c_2}(1)p_{c_1}(1)$;

$p_{b_2}(0)=p_{c_2}(0)\alpha_1(0)+p_{c_2}(1)\alpha_1(2)$;
$p_{b_2}(1)=p_{c_2}(1)\alpha_1(0)+p_{c_2}(0)\alpha_1(2)$;

and

$\hat{x}_{b_2}=p_{b_2}(0)-p_{b_2}(1)=(p_{c_2}(0)-p_{c_2}(1))(p_{c_1}(0)-p_{c_1}(1))=\dashbox{$\hat{x}_{c_2}\hat{x}_{c_1}$}$

(4) At time 3, we have

$\alpha_3(0)=p_{c_3}(0)p_{c_2}(0)$;
$\alpha_3(1)=p_{c_3}(0)p_{c_2}(1)$;
$\alpha_3(2)=p_{c_3}(1)p_{c_2}(0)$;
$\alpha_3(3)=p_{c_3}(1)p_{c_2}(1)$;

$p_{b_3}(0)=p_{c_3}(0)\alpha_2(0)+p_{c_3}(1)\alpha_2(1)+p_{c_3}(1)\alpha_2(2)+p_{c_3}(0)\alpha_2(3)$;

$p_{b_3}(1)=p_{c_3}(1)\alpha_2(0)+p_{c_3}(0)\alpha_2(1)+p_{c_3}(0)\alpha_2(2)+p_{c_3}(1)\alpha_2(3)$;

and

$\hat{x}_{b_3}=p_{b_3}(0)-p_{b_3}(1)=(p_{c_3}(0)-p_{c_3}(1))(p_{c_2}(0)-p_{c_2}(1))(p_{c_1}(0)-p_{c_1}(1))=\dashbox{$\hat{x}_{c_3}\hat{x}_{c_2}\hat{x}_{c_1}$}$

(5)Similarly we can have for any $k>=2$, we have

$\alpha_k(0)=p_{c_k}(0)p_{c_{k-1}}(0)$;
$\alpha_k(1)=p_{c_k}(0)p_{c_{k-1}}(1)$;
$\alpha_k(2)=p_{c_k}(1)p_{c_{k-1}}(0)$;
$\alpha_k(3)=p_{c_k}(1)p_{c_{k-1}}(1)$;

$p_{b_k}(0)=p_{c_k}(0)\alpha_{k-1}(0)+p_{c_k}(1)\alpha_{k-1}(1)+p_{c_k}(1)\alpha_{k-1}(2)+p_{c_k}(0)\alpha_{k-1}(3)$;

$p_{b_k}(1)=p_{c_k}(1)\alpha_{k-1}(0)+p_{c_k}(0)\alpha_{k-1}(1)+p_{c_k}(0)\alpha_{k-1}(2)+p_{c_k}(1)\alpha_{k-1}(3)$;

and
\begin{eqnarray}
\hat{x}_{b_k}&=&p_{b_k}(0)-p_{b_k}(1)=(p_{c_k}(0)-p_{c_k}(1))(\alpha_{k-1}(0)+\alpha_{k-1}(3)-\alpha_{k-1}(1)-\alpha_{k-1}(2))
\nonumber
\\ \nonumber &=& \dashbox{$\hat{x}_{c_k}\hat{x}_{c_{k-1}}\hat{x}_{c_{k-2}}$},
\end{eqnarray}
where

$(\alpha_{k-1}(0)+\alpha_{k-1}(3)-\alpha_{k-1}(1)-\alpha_{k-1}(2))=p_{c_{k-1}}(0)p_{c_{k-2}}(0)+p_{c_{k-1}}(1)p_{c_{k-2}}(1)-p_{c_{k-1}}(0)p_{c_{k-2}}(1)-p_{c_{k-1}}(1)p_{c_{k-2}}(0)=\hat{x}_{c_{k-1}}\hat{x}_{c_{k-2}}$.

Therefore, the decoder input and its output soft symbol estimates,
$\hat{x}_{c_k}$ and $\hat{x}_{b_k}$, for the code, generated by
$g_{FBC}(x)=\frac{1}{x^2+x+1}$, have the following relationship
\begin{equation} \label{eq3}
\hat{x}_{b_k}=\hat{x}_{c_k}\hat{x}_{c_{k-1}}\hat{x}_{c_{k-2}}.
\end{equation}

By taking the natural logarithm of both sides of the above equation,
we get
\begin{equation} \label{eq4}
ln\hat{x}_{b_k}=ln\hat{x}_{c_k}+ln\hat{x}_{c_{k-1}}+ln\hat{x}_{c_{k-2}}.
\end{equation}

\begin{figure}
\centering
\includegraphics[width=0.8\textwidth]{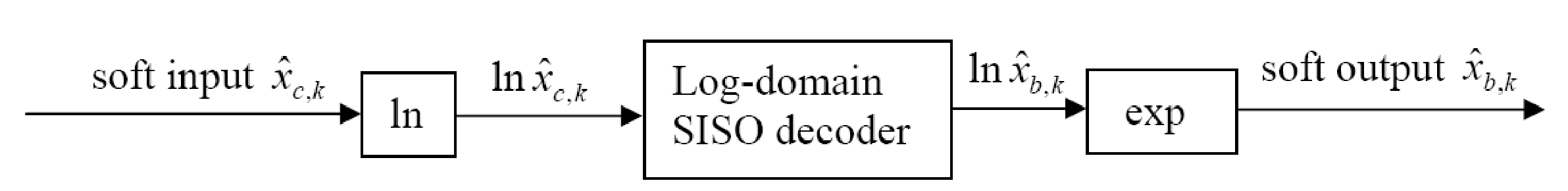}
\caption{The relationship of a SISO decoder and its Log-domain SISO
decoder} \label{fig1}
\end{figure}

We define the decoder with the input and output being the logarithm
of the soft symbol estimates (SSE) of the coded symbols and SSEs of
the information symbols, as the Log-domain soft-input-soft-output
(SISO) decoder. As shown in Fig. \ref{fig1}, the SISO decoder can be
implemented by adding a logarithm module and an exponential module
at the front and rear end of the log-domain SISO decoder,
respectively.

\begin{figure}
\centering
\includegraphics[width=0.6\textwidth]{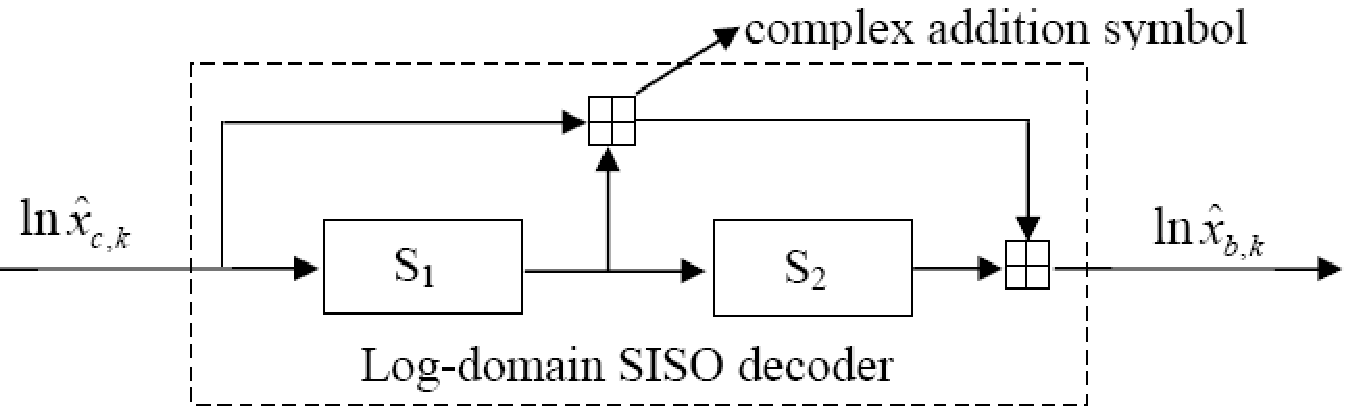}
\caption{The Log-domain SISO forward decoder implemented by using
its dual convolutional encoder} \label{fig2}
\end{figure}

Based on Eq. \ref{eq4}, log-domain SISO forward decoding of the code
$g_{FBC}(x)=\frac{1}{x^2+x+1}$ can be implemented by using the
convolutional encoder, generated by the generator polynomial
$1/g_{FBC}(x)=x^2+x+1$, as shown in Fig. \ref{fig2}. Here the
addition operation in the encoder is not carried out in the binary
field as in conventional convolutional encoders, but in the complex
field.

Eq. \ref{eq4} and Fig. \ref{fig2} reveal an interesting explicit
relationship of the binary encoder and SISO forward decoder of a
rate-1 feedback only convolutional code. This can be generalized to
any FBC codes as summarized in the following theorem.

 \textbf{Theorem 1 - Linear representation of forward decoding of a feedback only
convolutional (FBC) code:} For a FBC code, generated by a generator
polynomial $g_{FBC}(x)=1/q(x)$, we define its dual encoder as the
encoder with the inverse generator polynomial of $g_{FBC}(x)$, given
by $q_{FBC}(x)=1/g_{FBC}(x)=q(x)$. Then the log-domain SISO forward
decoding of the FBC code can be simply implemented by its dual
encoder in the complex field. This property is shown in Fig.
\ref{fig3}.

          Proof: See Appendix A.

\begin{figure}
\centering
\includegraphics[width=1\textwidth]{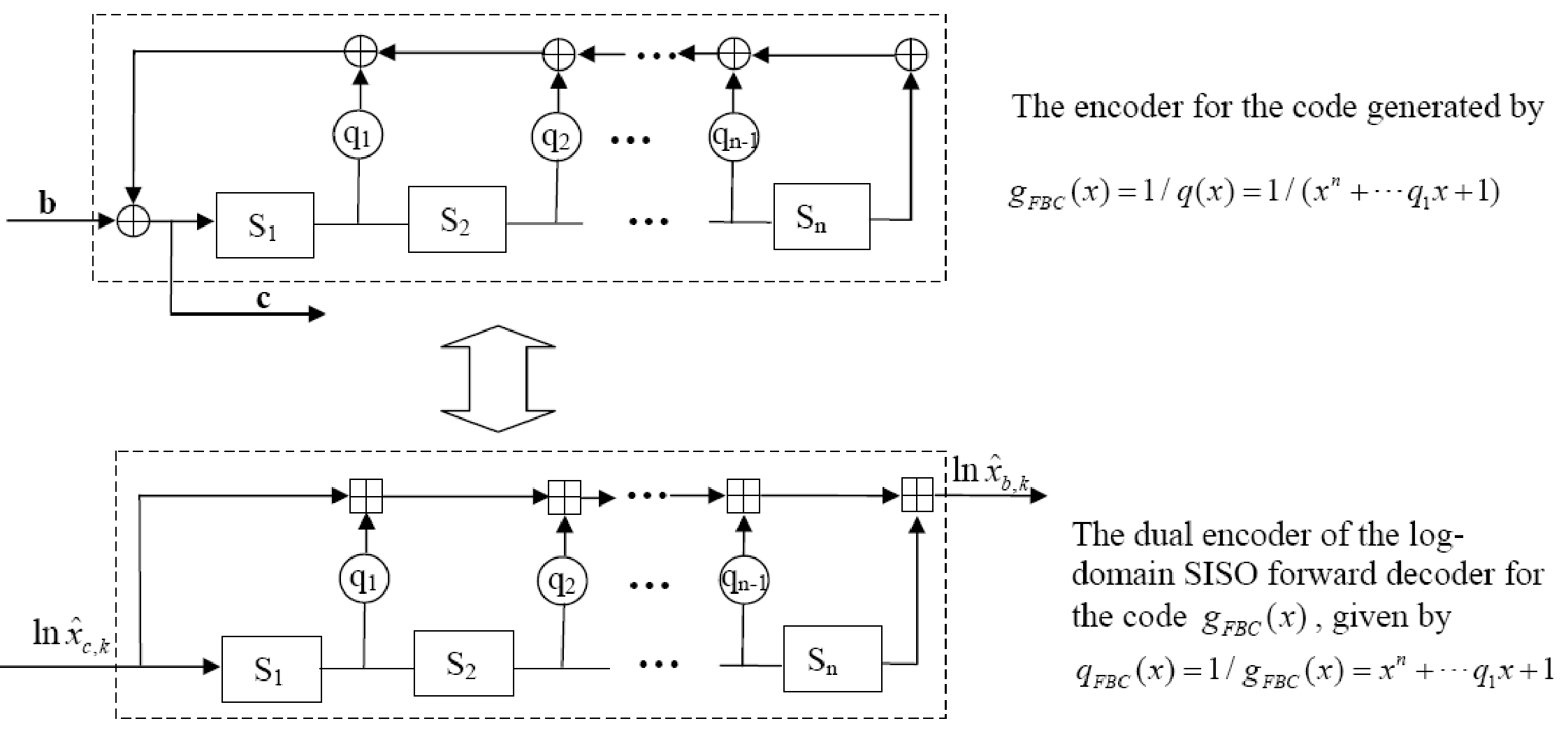}
\caption{Relationship of a FBC encoder and its Log-domain SISO
forward decoder} \label{fig3}
\end{figure}

\subsection{Forward decoding of feed-forward only convolutional (FFC)
code}

In this sub-section, we investigate the forward decoding of a FFC
code. As will be shown in the following example, the property shown
in Theorem 1 does not apply to such codes.

\begin{figure}
\centering
\includegraphics[width=0.8\textwidth]{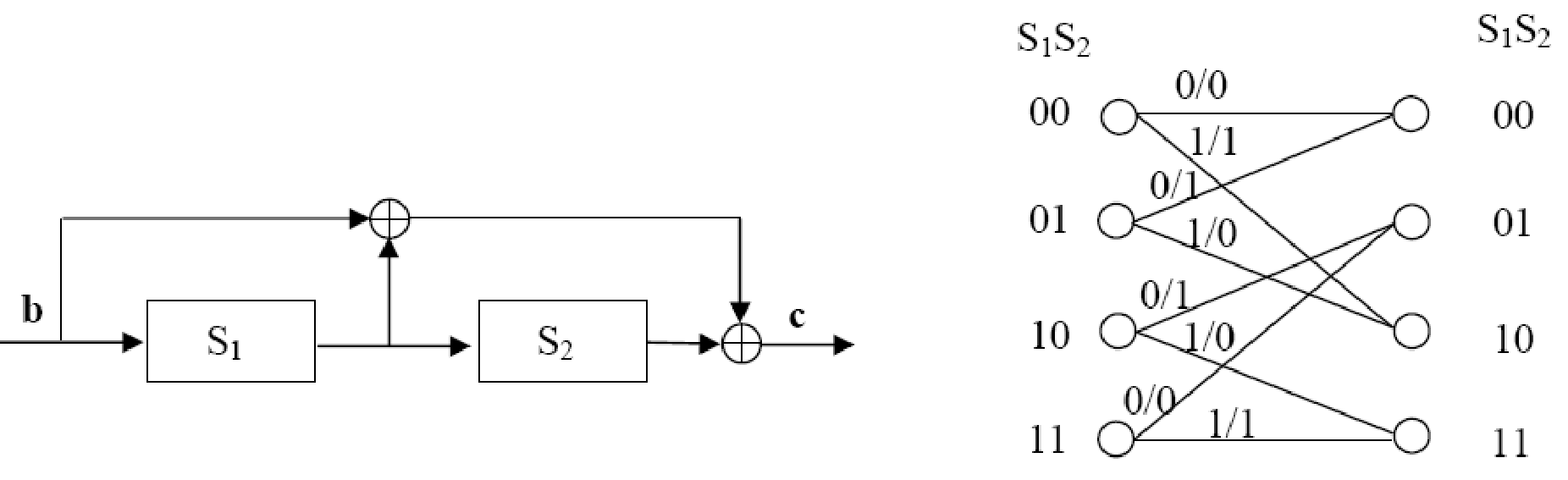}
\caption{The encoder and trellis of $g_{FFC}(x)=x^2+x+1$}
\label{fig3_0}
\end{figure}

  \textbf{Example 2}: We consider a FFC code with the generator polynomial of
$g_{FFC}(x)=x^2+x+1$ for which the trellis diagram and encoder are
shown in Fig. \ref{fig3_0}.

Let $ln\ddot{x}_{b_k}$ represent the output of the log-domain dual
encoder, generated based on Theorem 1, with the generator polynomial
of $q_{FFC}(x)=1/g_{FFC}(x)=1/(x^2+x+1)$. Table I compares
$ln\ddot{x}_{b_k}$ with the actual forward MAP decoding soft output
$ln\hat{x}_{b_k}$. Their differences are highlighted in the
dashed-line boxes.

\begin{table}
\caption{Comparison of the dual encoder output calculated based on
Theorem 1 $ln\ddot{x}_{b_k}$ with the actual forward MAP decoding
soft output $ln\hat{x}_{b_k}$}
\begin{tabular}{|c|c|c|c|c|}
\hline Log soft  & Memory $S_1$
of  & Memory $S_2$ of  & Log soft output of  & Desired soft decoding \\
input $ln\hat{x}_{c_k}$ & the dual encoder & the dual encoder & the dual encoder $ln\ddot{x}_{b_k}$ & output $ln\hat{x}_{b_k}$ \\
\hline
$ln\hat{x}_{c_1}$ & 0 & 0 & $ln\hat{x}_{c_1}$ & $ln\hat{x}_{c_1}$ \\
\hline $ln\hat{x}_{c_2}$ & $ln\hat{x}_{c_1}$ & 0 & $ln\hat{x}_{c_2}+ln\hat{x}_{c_1}$ & $ln\hat{x}_{c_2}+ln\hat{x}_{c_1}$ \\
\hline $ln\hat{x}_{c_3}$ & $ln\hat{x}_{c_2}+ln\hat{x}_{c_1}$ & $ln\hat{x}_{c_1}$ & $ln\hat{x}_{c_3}+ln\hat{x}_{c_2}$
+  & $ln\hat{x}_{c_3}+ln\hat{x}_{c_2}$\\
& & & \dashbox{$ln\hat{x}_{c_1}+ln\hat{x}_{c_1}$} & \\
\hline $ln\hat{x}_{c_4}$ & $ln\hat{x}_{c_3}$+$ln\hat{x}_{c_2}$  & $ln\hat{x}_{c_2}+ln\hat{x}_{c_1}$ & $ln\hat{x}_{c_4}$+$ln\hat{x}_{c_3}$ + $ln\hat{x}_{c_1}$+ & $ln\hat{x}_{c_4}$+$ln\hat{x}_{c_3}$ + $ln\hat{x}_{c_1}$\\
& + $ln\hat{x}_{c_1}$+$ln\hat{x}_{c_1}$ & & \dashbox{$ln\hat{x}_{c_2}$+$ln\hat{x}_{c_2}$+$ln\hat{x}_{c_1}$+$ln\hat{x}_{c_1}$} & \\
\hline $ln\hat{x}_{c_5}$ & $ln\hat{x}_{c_4}$+$ln\hat{x}_{c_3}$ + $ln\hat{x}_{c_1}$ & $ln\hat{x}_{c_3}$+$ln\hat{x}_{c_2}$  & $ln\hat{x}_{c_5}$+$ln\hat{x}_{c_4}$ + $ln\hat{x}_{c_2}$+$ln\hat{x}_{c_1}$ + & $ln\hat{x}_{c_5}$+$ln\hat{x}_{c_4}$ + $ln\hat{x}_{c_2}$+$ln\hat{x}_{c_1}$ \\
& +$ln\hat{x}_{c_2}$+$ln\hat{x}_{c_2}$+$ln\hat{x}_{c_1}$+$ln\hat{x}_{c_1}$ & + $ln\hat{x}_{c_1}$+$ln\hat{x}_{c_1}$ & \dashbox{$ln\hat{x}_{c_3}$ +$ln\hat{x}_{c_3}$+$ln\hat{x}_{c_2}$ +$ln\hat{x}_{c_2}$} & \\
& & &  \dashbox{+$ln\hat{x}_{c_1}$+$ln\hat{x}_{c_1}$+$ln\hat{x}_{c_1}$+$ln\hat{x}_{c_1}$} & \\
\hline \vdots & \vdots & \vdots & \vdots & \vdots \\
\hline

\end{tabular}\label{FONR_Table}
\end{table}

From the above table, we can see that the soft outputs of the dual
encoder, generated from Theorem 1, $ln\ddot{x}_{b_k}$  are different
from the actual forward MAP decoding soft outputs $ln\hat{x}_{b_k}$
when $k>2$. This is because the recursive structure of the dual
encoder $q_{FFC}(x)$ and the complex field addition operation of the
dual encoder. It can be observed from the above table that if the input
to the dual encoder is the binary symbol and addition in the encoder
is a module-2 addition, as in the conventional binary encoder, the
difference terms shown in the dotted-line-boxes will become zero and
the dual encoder output will be equal to the actual decoding output.
However, the inputs to the dual encoder are the logarithms of the
soft inputs, which are complex numbers, and the addition in the dual
encoder is done in the complex-number domain, which causes the
differences between  $ln\ddot{x}_{b_k}$ and $ln\hat{x}_{b_k}$. We
can observe from the table that the difference terms come from the
common terms of the shift-register contents $S_1$ and $S_2$ in the
dual encoder. If we can change structure of the dual encoder by
multiplying both the numerator and denominator by a common
polynomial, without changing its actual generator polynomial, such
that the encoder contents do not share any common elements at any
time instant, then the difference between $ln\ddot{x}_{b_k}$ and
$ln\hat{x}_{b_k}$ will disappear and the dual encoder output will be
equal to the actual MAP forward decoding output.

In Example 2, if we multiply both the numerator and denominator of
the dual encoder generator polynomial $q(x)$ by $(1+x)$, then we
have
\begin{equation} \label{eq5}
q(x)=\frac{1+x}{g_{FFC}(x)(1+x)}=\frac{1+x}{1+x^3}.
\end{equation}

\begin{figure}
\centering
\includegraphics[width=0.7\textwidth]{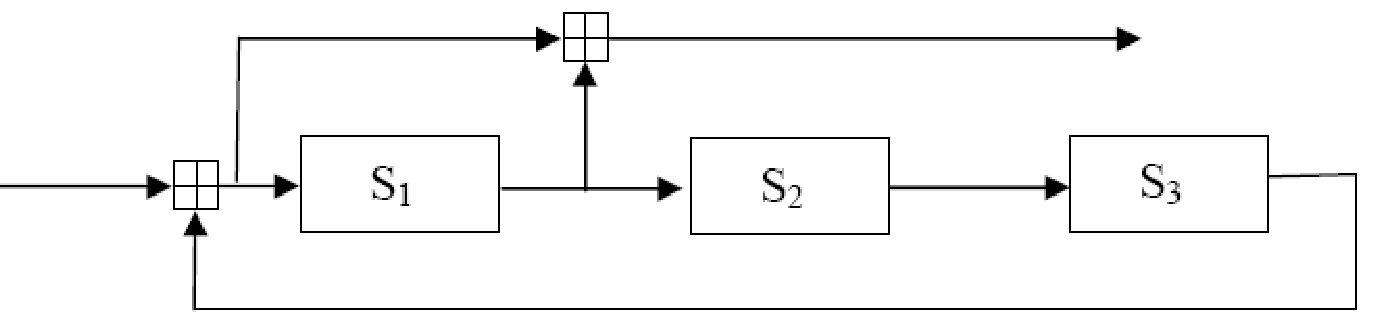}
\caption{The modified dual encoder of $g_{FFC}(x)=x^2+x+1$.}
\label{fig4}
\end{figure}

Fig. \ref{fig4} shows the encoder with the polynomial in Eq.
(\ref{eq5}).

Table \ref{FONR_Table} shows the outputs of the modified dual
decoder and the output of the actual MAP forward decoder. We can see
that the soft outputs of the modified dual encoder are exactly the
same as the actual MAP forward decoding outputs.

\begin{table}
\caption{Comparison of modified dual encoder output
$ln\ddot{x}_{b_k}$ with the actual forward MAP decoding soft output
$ln\hat{x}_{b_k}$.}
\begin{tabular}{|c|c|c|c|c|c|}
\hline Log soft   & Memory $S_1$ & Memory $S_2$ & Memory $S_3$ & Log soft output of   & Desired soft decoding \\
input $ln\hat{x}_{c_k}$ &  &   &   & the modified dual encoder $ln\ddot{x}_{b_k}$ & output $ln\hat{x}_{b_k}$\\

\hline
$ln\hat{x}_{c_1}$ & 0 & 0 & 0 & $ln\hat{x}_{c_1}$ & $ln\hat{x}_{c_1}$ \\
\hline $ln\hat{x}_{c_2}$ & $ln\hat{x}_{c_1}$ & 0 & 0 & $ln\hat{x}_{c_2}+ln\hat{x}_{c_1}$ & $ln\hat{x}_{c_2}+ln\hat{x}_{c_1}$ \\
\hline $ln\hat{x}_{c_3}$ & $ln\hat{x}_{c_2}$ & $ln\hat{x}_{c_1}$ & 0&  $ln\hat{x}_{c_3}+ln\hat{x}_{c_2}$ & $ln\hat{x}_{c_3}+ln\hat{x}_{c_2}$\\
\hline $ln\hat{x}_{c_4}$ & $ln\hat{x}_{c_3}$ & $ln\hat{x}_{c_2}$ & $ln\hat{x}_{c_1}$ & $ln\hat{x}_{c_4}$+$ln\hat{x}_{c_3}$ + $ln\hat{x}_{c_1}$ & $ln\hat{x}_{c_4}$+$ln\hat{x}_{c_3}$ + $ln\hat{x}_{c_1}$ \\
\hline $ln\hat{x}_{c_5}$ & $ln\hat{x}_{c_4}$+$ln\hat{x}_{c_1}$ & $ln\hat{x}_{c_3}$ & $ln\hat{x}_{c_2}$ & $ln\hat{x}_{c_5}$+$ln\hat{x}_{c_4}$ + $ln\hat{x}_{c_2}$+$ln\hat{x}_{c_1}$ & $ln\hat{x}_{c_5}$+$ln\hat{x}_{c_4}$ + $ln\hat{x}_{c_2}$+$ln\hat{x}_{c_1}$\\
\hline \vdots & \vdots & \vdots & \vdots & \vdots & \vdots \\
\hline

\end{tabular}\label{FONR_Table}
\end{table}

We can prove that for any FFC codes, we can always find a modified
dual decoder to implement a MAP forward decoder without changing its
actual generator polynomial. This is summarized in Theorem 2.

Before we present the new theorem, we first define a \textit{minimum
complementary polynomial}. For a given polynomial
$a(x)=x^n+\cdots+a_1x+1$, we define the \textit{minimum
complementary polynomial} as the polynomial of the smallest degree,
\begin{equation} \label{eq6}
z(x)=x^l+z_{l-1}x^{l-1}+\cdots+z_1x+1
\end{equation}
such that
\begin{equation} \label{eq7}
a(x)z(x)=x^{n+l}+1.
\end{equation}

Since $a(x)=x^n+\cdots+a_1x+1$ always divides $x^{2^n-1}+1$, the
minimum complementary polynomial of $a(x)$ always exists.

  \textbf{Theorem 2 - Linear presentation of forward decoding of a feed-forward only
convolutional (FFC) code:} For a FFC code, generated by a generator
polynomial $g_{FFC}(x)=a(x)$, let $z(x)$ represent its
\textit{minimum complementary polynomial} of degree $l$. The
log-domain SISO forward decoding of the FFC code can be implemented
by its dual encoder with the generator polynomial of
\begin{equation} \label{eq8}
q_{FFC}(x)=\frac{z(x)}{a(x)z(x)}=\frac{z(x)}{x^{n+l}+1}=\frac{x^l+z_{l-1}x^{l-1}+\cdots
+z_1x+1}{x^{n+l}+1}.
\end{equation}

   Proof: See Appendix B.
\\

As it can be noted from Theorem 2, in contrast to FBC, the encoder and decoder of which can be implemented by the same number of shift registers, for the FFC the number of shift registers required in decoder will be increased compared to the encoder and the number of increased shift registers depends on the degree of its minimum complementary polynomial.

Theorem 2 can be easily extended to a general convolutional (GC)
code as shown in the following corollary.

   \textbf{Corollary 1 - Linear presentation of forward decoding of a general
convolutional (GC) code}: For a GC code, generated by a generator
polynomial
$g_{GC}(x)=\frac{a(x)}{g(x)}=\frac{x^n+\cdots+a_1x+1}{x^n+\cdots+g_1x+1}$,
let $z(x)$ be the degree-$l$ minimum complementary polynomial of
$a(x)$. The log-domain SISO forward decoding of the GC code can be
simply implemented by its dual encoder with the generator polynomial
of
\begin{eqnarray} \label{eq9}
q_{GC}(x) &=&
\frac{g(x)z(x)}{a(x)z(x)}=\frac{g(x)z(x)}{x^{n+l}+1}=\frac{x^{n+l}+\cdots
+h_1x+1}{x^{n+l}+1} \\ \nonumber &=& 1+\frac{h_{n+l-1}x^{n+l-1}+
\cdots +h_1x}{x^{n+l}+1},
\end{eqnarray}
where $g(x)z(x)=x^{n+l}+h_{n+l-1}x^{n+l-1}+\cdots+h_1x+1$

\begin{figure}
\centering
\includegraphics[width=1\textwidth]{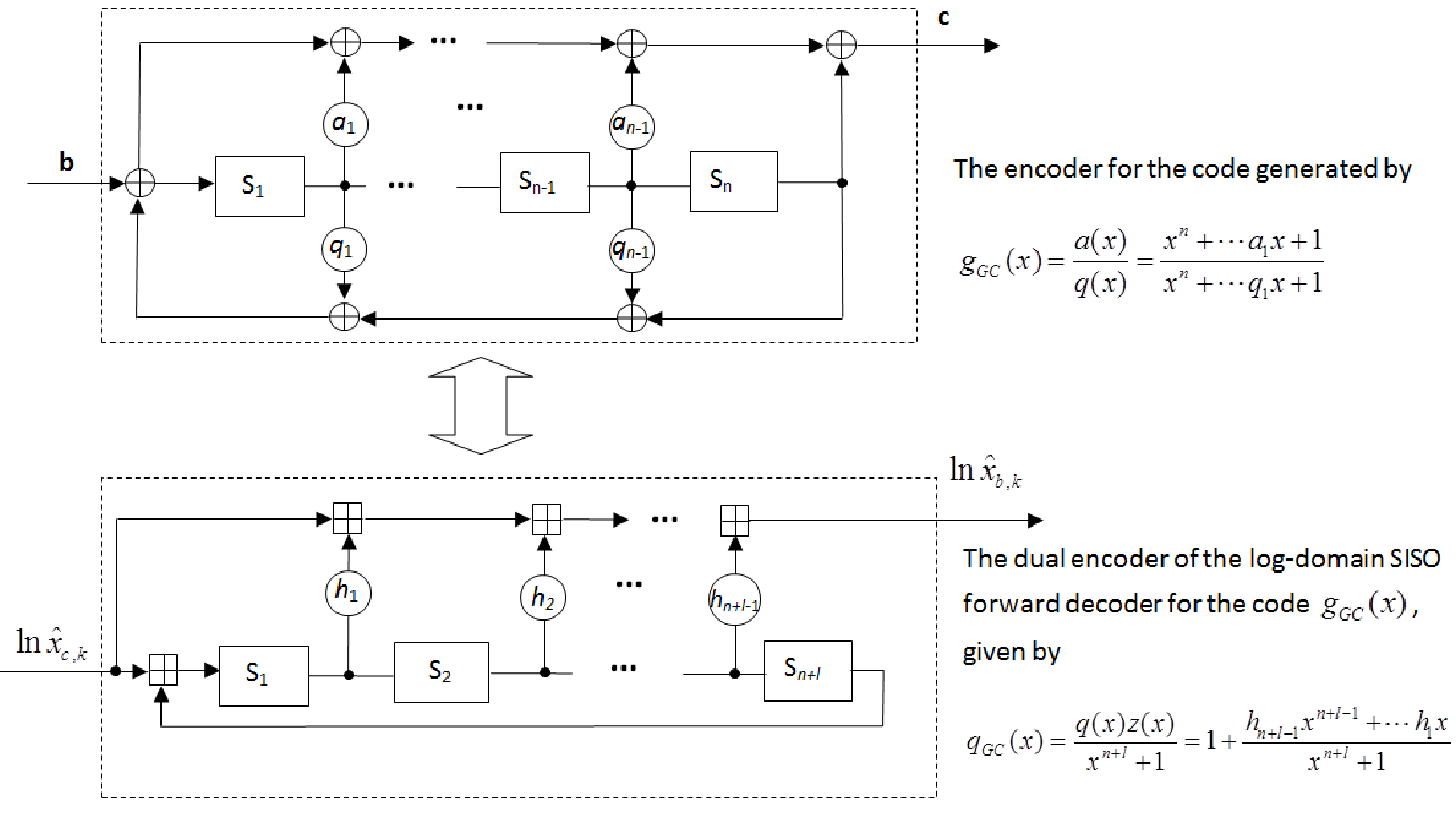}
\caption{The encoder and its dual encoder of forward decoding of a
general convolutional (GC) code} \label{fig5}
\end{figure}

This relationship of a binary encoder and its dual encoder is shown
in Fig. \ref{fig5}. Corollary 1 can be directly derived from Theorem
2, so we skip its proof here.

\section{Linear Presentation of Backward Decoding of Rate-1 Convolutional Codes}

\begin{figure}
 \centering
\subfigure[The encoder of $g(x)=a(x)/q(x)$]{
\includegraphics[width=0.8\textwidth]{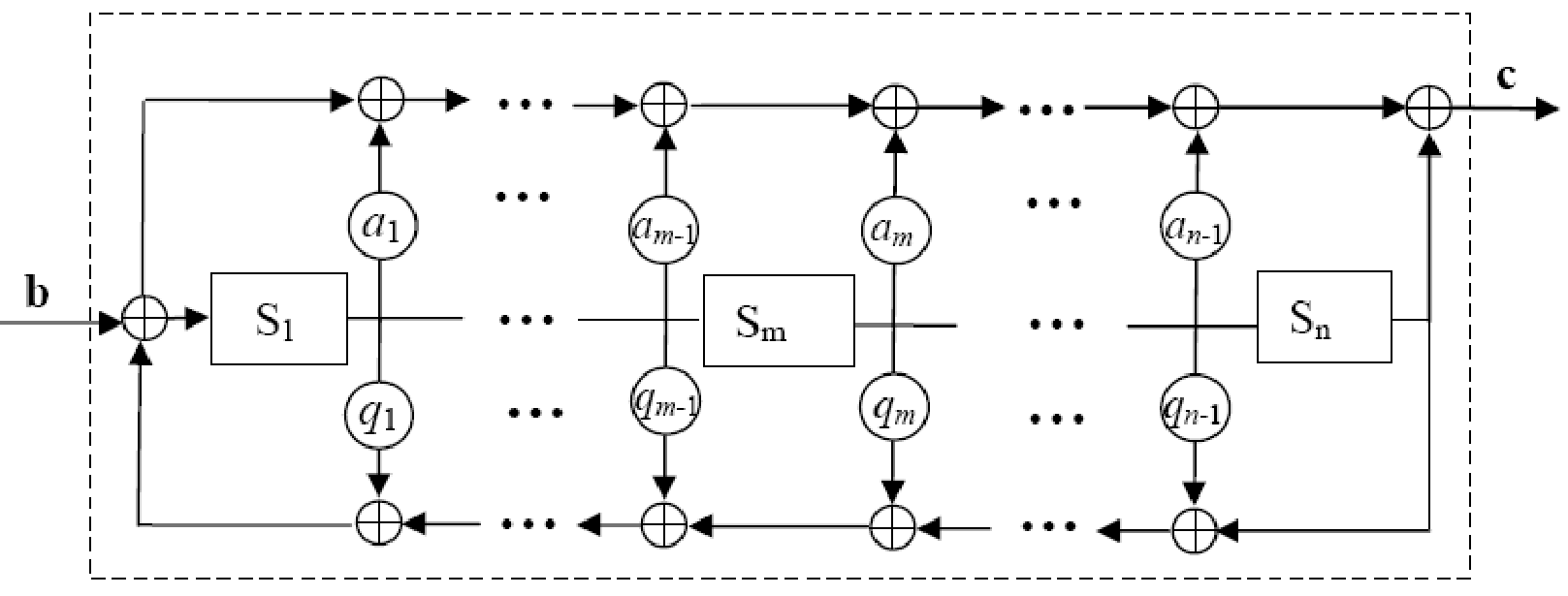}
\label{fig6_1} } \subfigure[The encoder of $g(x)=a(x)/q(x)$ with
reverse memory labeling]{
\includegraphics[width=0.8\textwidth]{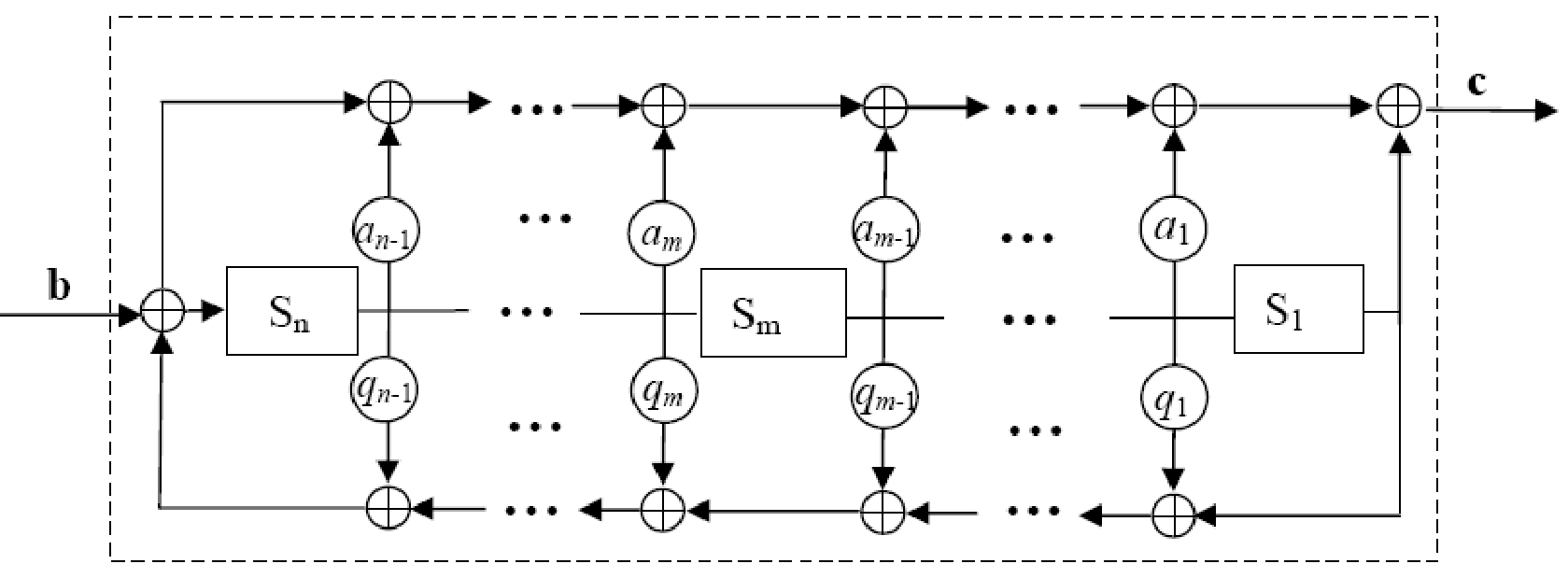}
\label{fig6_2}} \caption{An encoder with reverse memory labeling}
\end{figure}

In this section, we investigate the MAP backward decoding of rate-1
convolutional codes and derive its dual encoder structure. Before
discussing the backward decoding, we first define a reverse
memory-labeling of a general convolutional (GC) code. Given the
encoder of a GC code with rational generator polynomial
$g(x)=\frac{a(x)}{q(x)}=\frac{x^n+\cdots+a_1x+1}{x^n+\cdots+q_1x+1}$,
if we change the labeling of the $k$-th shift register in the
encoder from $S_k$ to $S_{n-k}$, and change their respective
feed-forward coefficient from $a_k$ to $a_{n-k}$,
$k$=$1,2,\ldots,n$, and feedback coefficients from $b_k$ to
$b_{n-k}$, $k$=$1,2,\ldots,n$, we will derive an encoder with a new
trellis. The resulting encoder is referred to as the \textit{reverse
memory-labeling encoder} of $g(x)$. Figs. \ref{fig6_1} and
\ref{fig6_2} show the encoder and the reverse memory-labeling
encoder of $g(x)$.

\begin{figure}
\centering \subfigure[The encoder of code
$g(x)=\frac{D^2+1}{D^2+D+1}$]{
\includegraphics[width=0.5\textwidth]{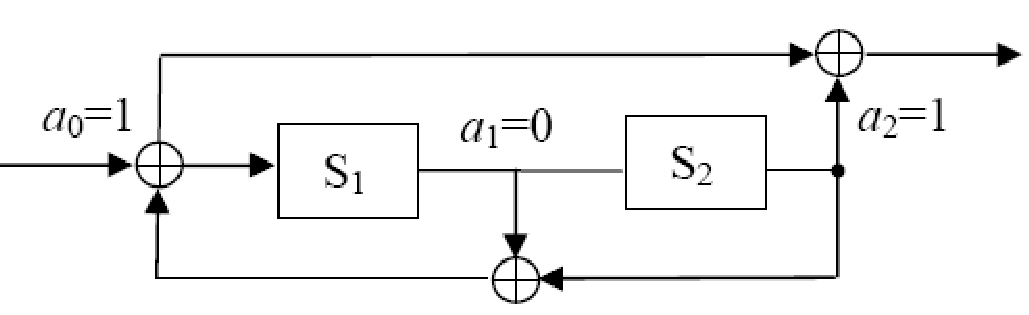}
\label{fig7_1}} \subfigure[The trellis of code $g(x)$]{
\includegraphics[width=0.3\textwidth]{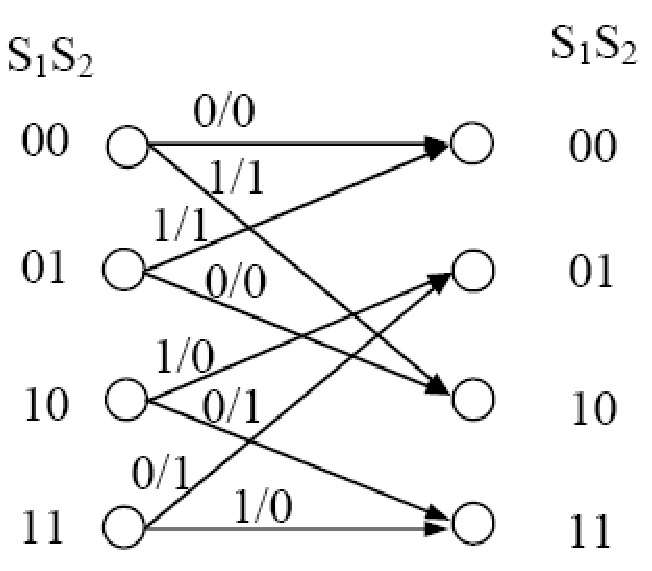}
\label{fig7_2}} \subfigure[The backward trellis of the code]{
\includegraphics[width=0.8\textwidth]{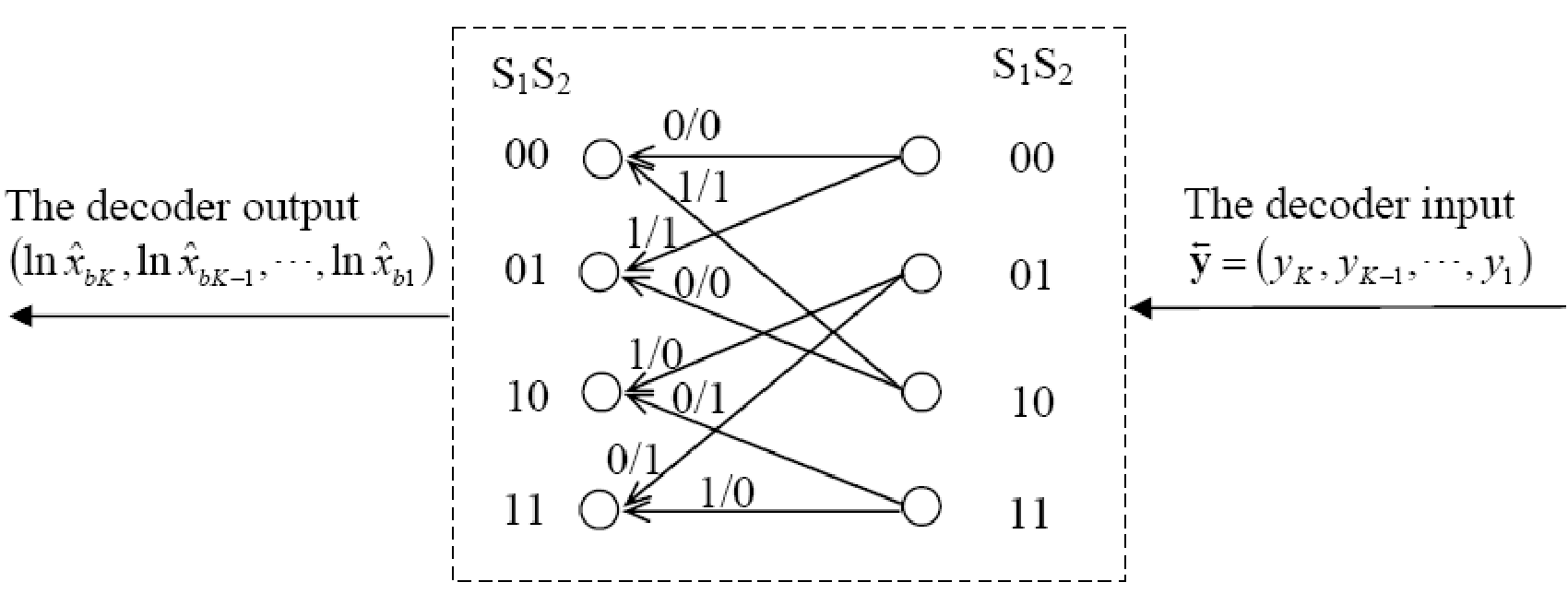}
\label{fig7_3}} \subfigure[The equivalent forward representation of
backward trellis]{
\includegraphics[width=0.8\textwidth]{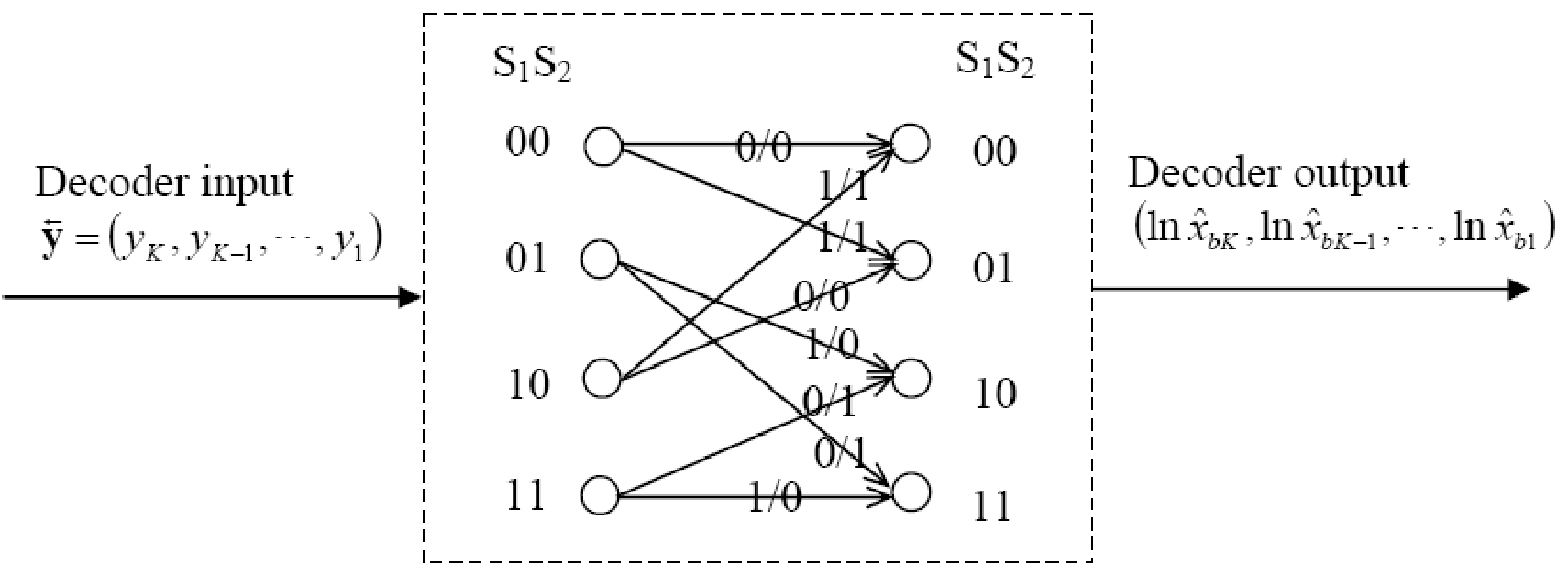}
\label{fig7_4}} \subfigure[The encoder corresponds to the trellis of
($d$)]{
\includegraphics[width=0.5\textwidth]{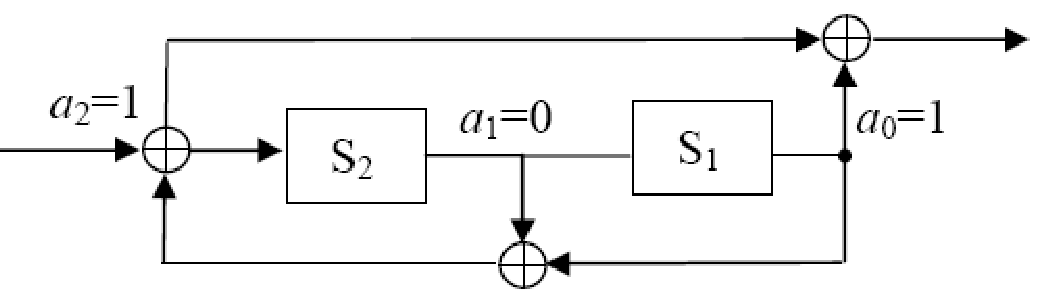}
\label{fig7_5}} \caption{Trellis, backward trellis and their
respective encoders for the code $g(x)=\frac{D^2+1}{D^2+D+1}$}
\end{figure}

In a MAP backward decoding, the received signals are decoded
backward in a time-reverse order. That is, given the received signal
sequence $\mathbf{y}=(y_1,y_2,\ldots,y_K)$, the order of signals to
be decoded is from $y_K$, $y_{K-1}$, till $y_1$.  In order to decode
the received signals backward, the decoder has to follow the trellis
in a reverse direction. Figs. \ref{fig7_1} and \ref{fig7_2} show the
encoder and trellis of the code with the generator polynomial
$g(x)=\frac{D^2+1}{D^2+D+1}$. Fig. \ref{fig7_3} shows the backward
trellis. For the decoder with the backward trellis in Fig.
\ref{fig7_3}, the input to the decoder is at the right hand side of
the decoder and its output is at the left hand side, which operates
in a reverse direction of the conventional decoder. Fig.
\ref{fig7_4} shows the corresponding forward representation of the
backward trellis, where the decoder input and output are changed to
the conventional order. The forward representation of the backward
trellis can be implemented by an encoder shown in Fig. \ref{fig7_5}.
When we compare Figs. \ref{fig7_1} and \ref{fig7_5}, it can be
easily seen that the encoder in Fig. \ref{fig7_1} is the encoder of
code $g(x)=\frac{D^2+1}{D^2+D+1}$ and that in Fig. \ref{fig7_5} is
its encoder with the \textit{reverse memory-labeling}.

This relationship of the encoders for the forward and backward
trellises can be extended to general rate-1 convolutional codes, as
shown in the following theorem.

  \textbf{Theorem 3}: Given an encoder with a generator polynomial
  $g(x)=\frac{a(x)}{q(x)}=\frac{x^n+\cdots+a_1x+1}{x^n+\cdots+q_1x+1}$, the forward representation of its backward trellis can be
implemented by its \textit{reverse memory-labeling encoder} of the
same generator polynomial $g(x)$.

  Proof: See Appendix C.
\\

From Theorem 2, we know that the log-domain SISO forward decoding of
a given general convolutional (GC) encoder with a generator
polynomial $g(x)=\frac{a(x)}{q(x)}$ can be implemented by its dual
encoder with the generator polynomial
$q_{GC}(x)=\frac{q(x)z(x)}{a(x)z(x)}$, where $z(x)$ is the
degree-$l$ minimum complementary polynomial of $a(x)$. Then
according to Theorem 3, the log-domain SISO backward decoding of the
GC code can be implemented by the\textit{ reverse memory-labeling
encoder} of $q_{GC}(x)$. By combining Theorems 2 and 3, we can
obtain the linear presentation of backward decoding, which is
summarized in the following Theorem.

\begin{figure}
\centering
\includegraphics[width=1\textwidth]{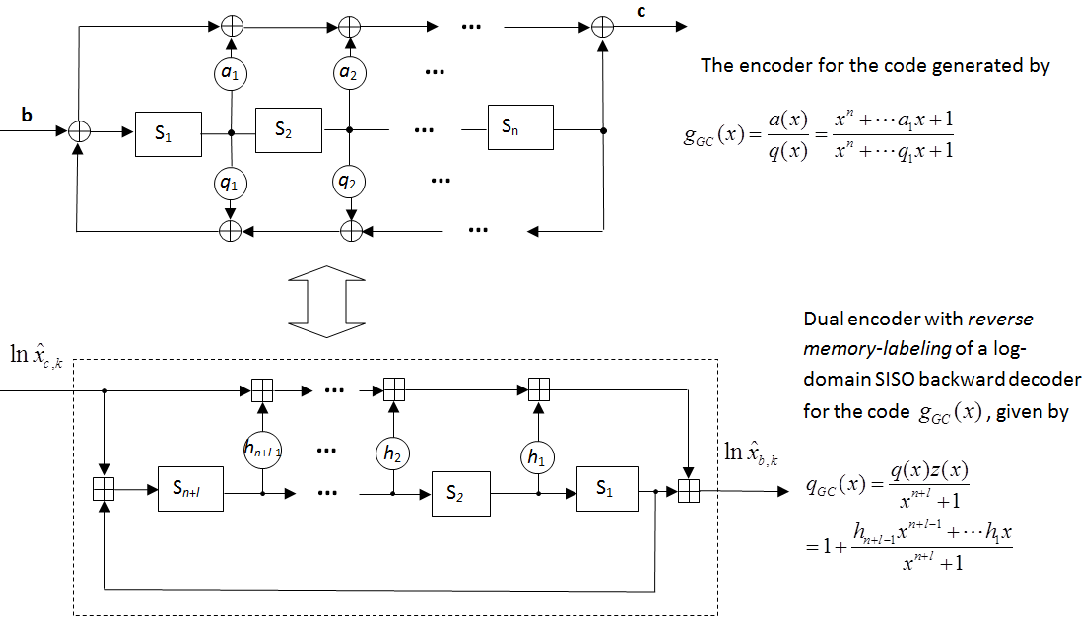}
\caption{The encoder and its dual encoder for backward decoding of a
general convolutional code} \label{fig8}
\end{figure}

  \textbf{Theorem 4 - Linear presentation of backward decoding of a general
convolutional (GC) code:} We consider a general convolutional
encoder with a generator polynomial of
$g(x)=\frac{a(x)}{q(x)}=\frac{x^n+\cdots+a_1x+1}{x^n+\cdots+q_1x+1}$.
Let $z(x)$ be the degree-$l$ minimum complementary polynomial of
$a(x)$. Its log-domain SISO backward decoding can be implemented by
its dual encoder with \textit{reverse memory-labeling} and the
generator polynomial of
\begin{eqnarray} \label{eq10}
q_{GC}(x) &=&
\frac{q(x)z(x)}{a(x)z(x)}=\frac{q(x)z(x)}{x^{n+l}+1}=\frac{x^{n+l}+\cdots
+h_1x+1}{x^{n+l}+1}  \\ \nonumber &=&
1+\frac{h_{n+l-1}x^{n+l-1}+\cdots +h_1x}{x^{n+l}+1}.
\end{eqnarray}
This presentation is shown in Fig. \ref{fig8}.

From Theorem 4, we can easily derive the backward decoding
presentation of a feed-forward only convolutional (FFC) code,
summarized in the following Corollary.

   \textbf{Corollary 2 - Linear presentation of backward decoding of a feed-forward
only convolutional (FFC) code:} For a FFC code, generated by a
generator polynomial $g_{FFC}(x)=a(x)=x^n+\ldots+a_1x+1$, let $z(x)$
be the degree-$l$ minimum complementary polynomial of $a(x)$. Its
log-domain SISO backward decoding can be implemented by its dual
encoder with \textit{reverse memory-labeling} and generator
polynomial
\begin{equation} \label{eq11}
q_{FFC}(x)=\frac{z(x)}{a(x)z(x)}=\frac{z(x)}{x^{n+l}+1}=\frac{x^{l}+z_{l-1}x^{l-1}\cdots
+z_1x+1}{x^{n+l}+1}.
\end{equation}

Corollary 2 can be proved in the same way as Theorem 4, so we skip
the proof here.

For a feedback only convolutional (FBC) code, we can prove that
backward decoding does not contribute to the MAP calculation. The
BCJR MAP decoding is exactly the same as the forward decoding. This
is summarized in the following Theorem.

  \textbf{Theorem 5 - Linear presentation of decoding of a feedback only
  convolutional (FBC) code:} For a FBC code, generated by a generator polynomial
$g_{FBC}(x)=1/q(x)$, the MAP forward decoding is in fact equivalent
to the BCJR MAP decoding. Its log-domain SISO decoder can be simply
implemented by the dual encoder of the MAP forward decoding with the
inverse generator polynomial of $g_{FBC}(x)$, given by
$q_{FBC}(x)=1/g_{FBC}(x)$.

  Proof: See Appendix D.
\\

From Theorem 5, we can see that the MAP decoder of a FBC code can be
implemented by its dual encoder using shift registers. This
significantly reduces the decoding complexity.

\section{The Representation of Bidirectional MAP Decoding}

In the previous two sections, we have introduced the linear
presentation of SISO MAP forward/backward decoding. Based on the
derived linear presentation, in this section, we represent the
bidirectional MAP decoder by linearly combining shift register
contents of the dual encoders of the respective forward and backward
decoders. We prove that such linear combining produces exactly the
same decoding output as the bidirectional BCJR MAP decoding.

Next, let us first discuss the boundary conditions for the dual
encoder. That is, how to determine the tail bits for the dual
encoder such that the state of dual encoder returns to all-zero
state at the end of encoding process. As we will discuss shortly,
the boundary conditions are essential for shift register contents
combining in the proposed decoding structure using dual encoders.

\subsection{Boundary conditions}\label{boundary consistency}

Let us consider a binary encoder $\bar{C}$ of memory length $n+l$,
described by $q_{GC}(x)=1+\frac{h_1x+ \cdots +
h_{n+l-1}x^{n+l-1}}{1+x^{n+l}}$. It has the same generator
polynomial as the dual encoder of a GC code C, generated by
$g_{GC}(x)=\frac{a(x)}{q(x)}$. Therefore, if the input to the
encoder $\bar{C}$ is a codeword ${\bf c}=\left(c_1, c_2, \cdots,
c_K\right)$, generated by $g_{GC}(x)$, the output of the encoder
$\bar{C}$ will produce the decoded binary information sequence $\bf
b$. Let us define $\left(c_{K+1},...,c_{K+n+l}\right)$ as the tail
bits required to terminate the encoder $\bar{C}$ at the all-zero
state. Then the tail biting convolutional encoder $\bar{C}$ has the
following property.

\newtheorem{lemma}{\bf Lemma}
\begin{lemma}\label{dual decoder and encoder states returen to zero}
The tail bits that terminate the encoder $\bar{C}$, described by
$q_{GC}(x)=1+\frac{h_1x+ \cdots + h_{n+l-1}x^{n+l-1}}{1+x^{n+l}}$,
at the all-zero state also terminate the encoder C, generated by
$g_{GC}(x)=\frac{a(x)}{q(x)}$, at the all-zero state.
\end{lemma}

\begin{proof}
See Appendix E.
\end{proof}

\begin{lemma}\label{same state transitions}
For a tail biting convolutional encoder $\bar{C}$, generated by
$q_{GC}(x)$, and a given input sequence $(c_1, c_2, \cdots, c_K,
c_{K+1}, \cdots, c_{K+n+l})$,  we define its backward encoder as the
encoder of the same generator polynomial with reverse-memory
labeling and time-reverse input $\left(c_{K+n+l}, \cdots , c_{K+1},
c_K, \cdots, c_2, c_1\right)$. Then the tail biting encoder
$\bar{C}$ and its backward encoder arrive at the same state at any
time $k$.
\end{lemma}

\begin{proof}
See Appendix F.
\end{proof}

\subsection{Shift register contents of the dual encoders for forward and backward decoding}
In the decoding structures we introduced in the previous two
sections, the input, output and shift register contents of dual
encoders for forward and backward decoding are all soft symbol
estimates (SSE). Let us consider a GC code, generated by
$g(x)=\frac{a(x)}{q(x)}$. Let $\overrightarrow{\hat{V}_j}(k)$ and
$\overleftarrow{\hat{V}_j} (k)$, $j=1, 2, \cdots, n+l$, $k=1, 2,
\cdots, K+n+l$, represent the $j$th shift register content of the
dual encoders for forward and backward decoding at time $k$,
described by the polynomial $q_{GC}(x)$.


To derive the bidirectional soft decoder output, we combine the
shift register contents of dual encoders for forward and backward
decoding in an optimal way. Let $\overrightarrow{S'_{i}}(k)$ and
$\overleftarrow{S'_{i}}(k), i=1, 2, \cdots, n+l,$ denote the memory
of the $i$th shift register at time $k$ in the encoder $\bar{C}$ and
its backward encoder. Let $P_{\overrightarrow{S'_{i}}(k)}(\omega)$
and $P_{\overleftarrow{S'_{i}}(k)}(\omega)$ denote the probability
of $\overrightarrow{S'_{i}}(k)=\omega$ and
$\overleftarrow{S'_{i}}(k)=\omega$ in the dual encoders of forward
and backward decoding, respectively. Their corresponding LLRs are
denoted by $\overrightarrow{L}_{S'_i}(k)$ and
$\overleftarrow{L}_{S'_i}(k)$. Their combined LLR is denoted by
$L_{S'_i}(k)$. Since $\overrightarrow{L}_{S'_i}(k)$ and
$\overleftarrow{L}_{S'_i}(k)$ are obtained from the forward decoding
based on the received signals from time 1 to $k$ and that from
backward decoding based on the received signals from time $K+n+l$ to
$k+1$, they are independent. Furthermore, as shown in Lemma
\ref{same state transitions}, for tail bitting encoder $\bar{C}$,
generated by $q_{GC}(x)$, forward and backward encoders will arrive
at the same state at time $k$. Therefore, in the optimal combining,
we have
\begin{align}\label{relationship of combined LLR and forward and backward LLR}
L_{S'_i}(k)=\overrightarrow{L}_{S'_i}(k)
+\overleftarrow{L}_{S'_i}(k).
\end{align}

Converted into the SSE representation, (\ref{relationship of
combined LLR and forward and backward LLR}) can be rewritten as
\begin{align}
\hat{V}_j(k)=\frac{\overrightarrow{\hat{V_{j}}}(k)+\overleftarrow{\hat{V_{j}}}(k)}{1+\overrightarrow{\hat{V_{j}}}(k)\overleftarrow{\hat{V_{j}}}(k)}.
\end{align}

Based on the dual encoder structure in Fig. 7b, the bidirectional
soft decoder output can be obtained from the combined shift register
contents as
\begin{align}\label{shift register combined output of the dual encoder}
\ln \hat{x}_{b_k}= \ln \hat{x}_{c_k}+\sum_{i=1}^{n+l-1} h_i \ln
\hat{V}_{i}(k-1).
\end{align}

As shown in the following theorem, such combining will produce
exactly the same output as the bidirectional BCJR MAP algorithm.

   \textbf{Theorem 6 - Shift register content combining of dual encoders of forward and backward decoding}: We can represent the bidirectional MAP decoder by linearly combining shift register contents of the dual encoders for forward and
backward decoding, as shown in (\ref{shift register combined output
of the dual encoder}). Such a combining produces exactly the same
decoding output as the bidirectional BCJR MAP decoding.

\begin{proof}
See Appendix G.
\end{proof}

\section{Simulation Results}

In this section, we provide the simulation results. All simulations
are performed for the BPSK modulation and a frame size of $K$=128
symbols over AWGN channels.

\begin{figure}[!htbp]
\centering
\begin{minipage}[t]{0.48\textwidth}
    \centering
    \makeatletter
    \def\fnum@figure{Fig. 11}
    \makeatother
    \renewcommand{\captionlabeldelim}{}
    \includegraphics[width=1\textwidth]{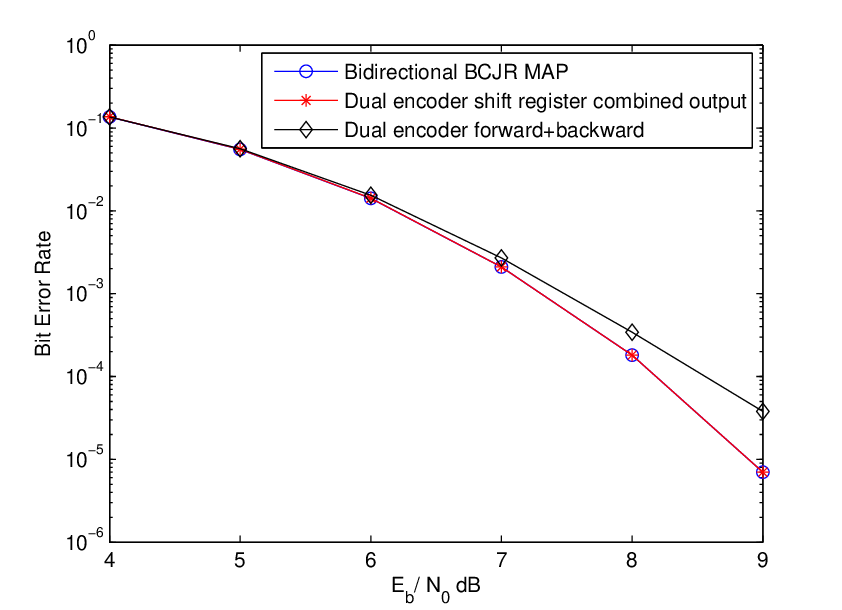}
    \caption{BER performances of code [5/7]$_8$}
    \label{57_GR_sim}
\end{minipage}
\begin{minipage}[t]{0.48\textwidth}
    \centering
    \makeatletter
    \def\fnum@figure{Fig. 12}
    \makeatother
    \renewcommand{\captionlabeldelim}{}
    \includegraphics[width=1\textwidth]{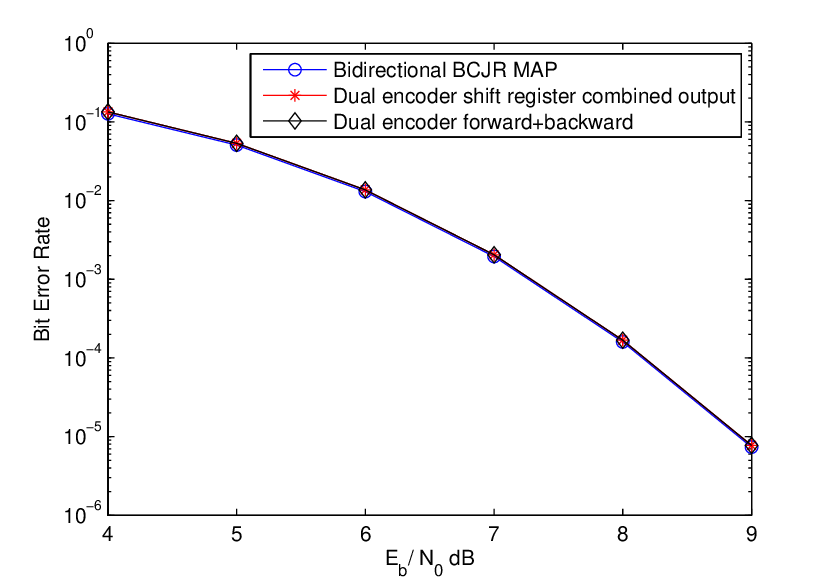}
    \caption{BER performances of code [5]$_8$}
    \label{5_FONR_sim}
\end{minipage}
\end{figure}

\begin{figure}[!htbp]
\centering
\begin{minipage}[t]{0.48\textwidth}
    \centering
    \makeatletter
    \def\fnum@figure{Fig. 13}
    \includegraphics[width=1\textwidth]{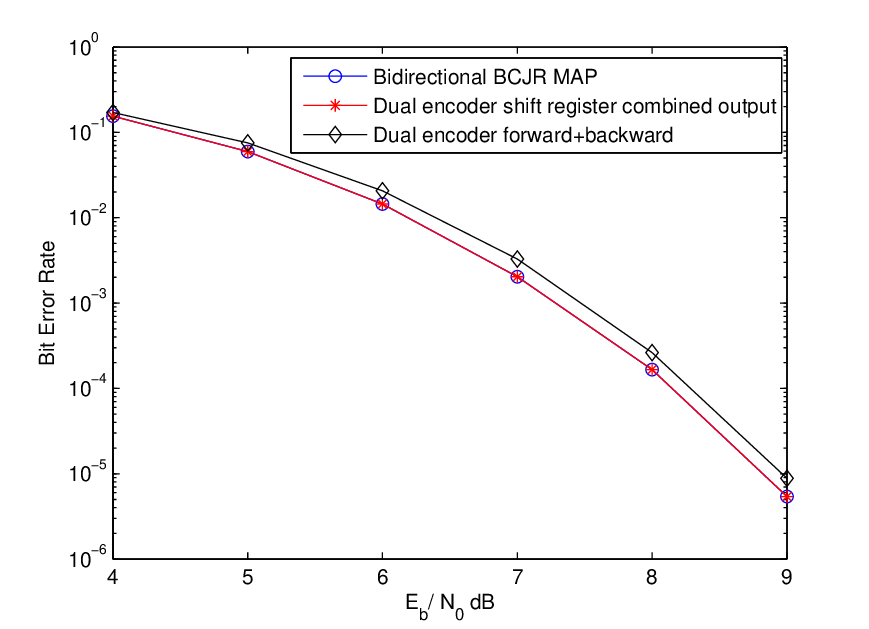}
\caption{BER performances of code [7]$_8$} \label{7_FONR_sim}
\end{minipage}
\begin{minipage}[t]{0.48\textwidth}
    \centering
    \makeatletter
    \def\fnum@figure{Fig. 14}
   \includegraphics[width=1\textwidth]{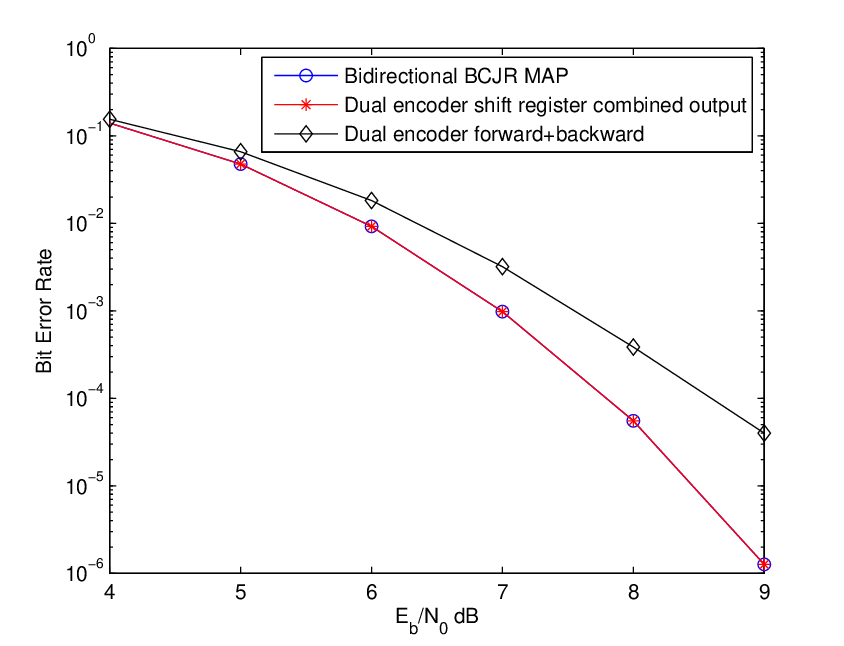}
\caption{BER performances of code [15/13]$_8$} \label{1513_GR_sim}
\end{minipage}
\end{figure}

\begin{figure}[!htbp]
\centering
\begin{minipage}[t]{0.48\textwidth}
    \centering
    \makeatletter
    \def\fnum@figure{Fig. 15}
\includegraphics[width=1\textwidth]{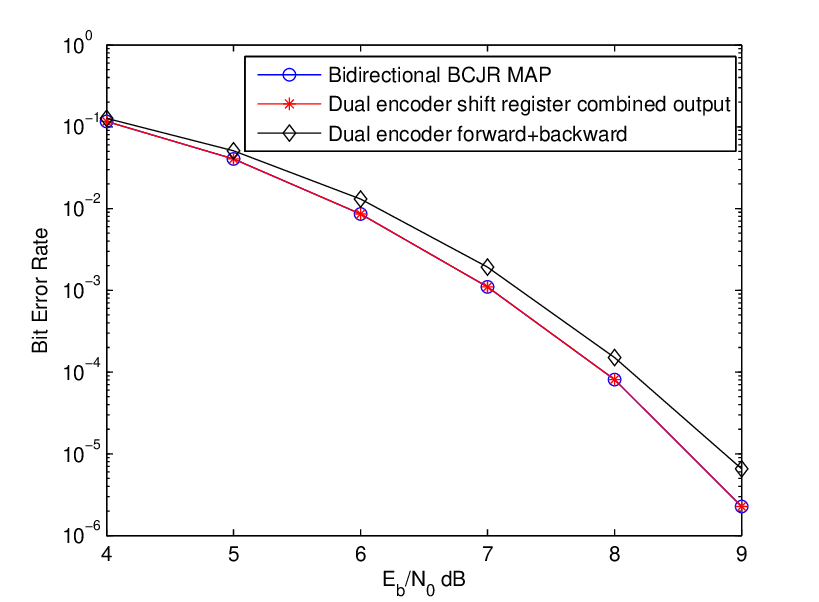}
\caption{BER performances of  code [17]$_8$} \label{17_FONR_sim}
\end{minipage}
\begin{minipage}[t]{0.48\textwidth}
    \centering
    \makeatletter
    \def\fnum@figure{Fig. 16}
\includegraphics[width=1\textwidth]{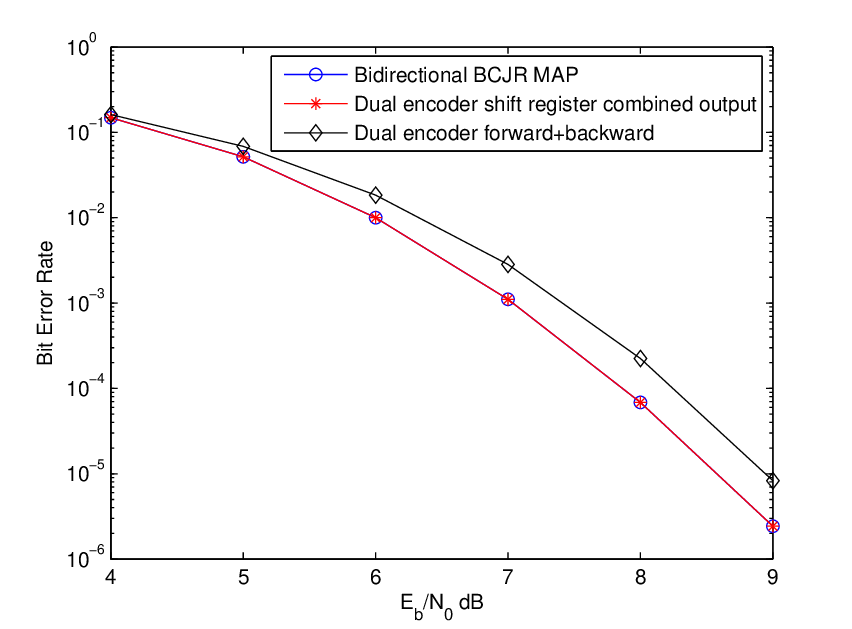}
\caption{BER performances of code [15]$_8$} \label{15_FONR_sim}
\end{minipage}
\end{figure}

Figs. \ref{57_GR_sim} to \ref{15_FONR_sim} show the bit error rate
(BER) performance of various 4-state and 8-state GC and FFC codes,
where the curve 'Dual encoder forward + backward' refers to the
direct summation of forward and backward dual encoder outputs, and
the curve 'Dual encoder shift register combined output' refers to
the optimal combining of forward and backward dual encoders as shown
in Theorem 6.

From figures, we can see that direct summation of forward and
backward dual encoder outputs has about $1dB$ performance loss when
compared to the bidirectional MAP decoding for the GC code [5/7]$_8$
at the BER of $10^{-5}$. This performance loss is reduced to around
$0.2dB$, $0.3dB$, and $0.4dB$ for $[7]_8$ FFC, $[17]_8$ FFC, and
$[15]_8$ FFC codes, respectively and increased to more than $1dB$
for the $[15/13]_8$ GC code. However, when we apply the shift
register combining approach detailed in Section IV to the forward
and backward dual encoder, their performance is exactly the same as
the BCJR MAP decoding. One particular point needs to be noted is
that for the FFC code [5]$_8$ the direct summation of forward and
backward dual encoder outputs has the same performance as the MAP
decoding, so no linear combination is actually required.

\section{Conclusions}

In this paper, we revisited the MAP forward and backward decoding
process for the rate-1 convolutional codes. Dual encoder structures
of forward and backward decoding for three different classes of
rate-1 convolutional codes are derived. The input to the dual
encoder is the logarithm of soft symbol estimates of the coded
symbols obtained from the received signals, and the dual encoder
output produces the logarithm of the soft symbol estimates of the
information symbols. For the general convolutional (GC) codes,
generated by a generator polynomial $g_{GC}(x)=\frac{a(x)}{q(x)}$,
the forward and backward decoding can be implemented by their
corresponding dual encoders, which are generated by the polynomial,
$\frac{q(x)z(x)}{a(x)z(x)}$, where $z(x)$ is the \textit{minimum
complementary polynomial} of $a(x)$. The feed-forward only
convolutional (FFC) code is just a special case of GC code, so it
has the same dual encoder structures as the GC code. The derived
linear presentation of decoder significantly reduced the the
computational complexity of MAP forward and backward recursion from
exponential to linear. Similarly, the bidirectional MAP decoder of
GC and FFC codes can be implemented by linearly combining the shift
register contents of dual encoders for the forward and backward
decoding. For a feedback only convolutional (FBC) code
$g_{FBC}(x)=\frac{1}{q(x)}$, the bidirectional MAP SISO decoder is
equivalent to the dual encoder for the forward decoding, with the
generator polynomial $q(x)$.

In this paper, we have only focused on a class of convolutional
codes, named rate-1 binary code. It is significant as component
codes in concatenated coding schemes, such as turbo coding. Also,
the linear presentation of MAP decoding derived in this paper can
also be applied to other codes and other applications. For example,
the transmission of digital signals in the presence of inter-symbol
interference (ISI) can also be represented by a convolutional
encoding process. The channel transfer function of an ISI channel
can be represented by a rate-1 convolutional encoder. Thus the
linear presentation of decoding can also be applied to facilitate
the MAP channel detection in ISI channels. Similarly, these
properties should exist for other linear codes that are amenable to
representation by a trellis diagram. We will discuss these in the
next series papers.

\section{Appendix}

\subsection{Proof of Theorem 1}

Let us consider a feedback only convolutional (FBC) code, generated
by a generator polynomial
\begin{equation} \label{eq12}
{g_{FBC}}(x) = 1/q(x) = 1/({q_n}{x^n} +  \cdots {q_1}x + 1),
\end{equation}
its encoder is shown in Fig. \ref{fig3}. Let ${S_i}(k),i = 1, \ldots
,n$  represent the state of memory $i$ at time $k$. Then according
to Fig. \ref{fig3} we have
\begin{equation} \label{eq13}
{c_k} = {S_1}(k) = {b_k} \oplus \sum\limits_{i = 1}^n {{q_i}{S_i}(k
- 1)}
\end{equation}
\begin{equation} \label{eq14}
{S_p}(k) = {S_{p - 1}}(k - 1),p\geq 2,
\end{equation}
where all summations are done in GF(2).

We can rewrite the above equation as follows
\begin{equation} \label{eq15}
{b_k} = {c_k} \oplus \sum\limits_{i = 1}^n {{q_i}{S_i}(k - 1)}
\end{equation}
\begin{equation} \label{eq16}
{S_1}(k) = {c_k}, {S_p}(k) = {S_{p - 1}}(k - 1), p\geq 2
\end{equation}
\begin{equation} \label{eq17}
{S_p}(k) = {c_{k - p}},
\end{equation}
where we assume that $c_k=0$ for $k\leq 0$.

\begin{figure}
\centering
\includegraphics[width=0.7\textwidth]{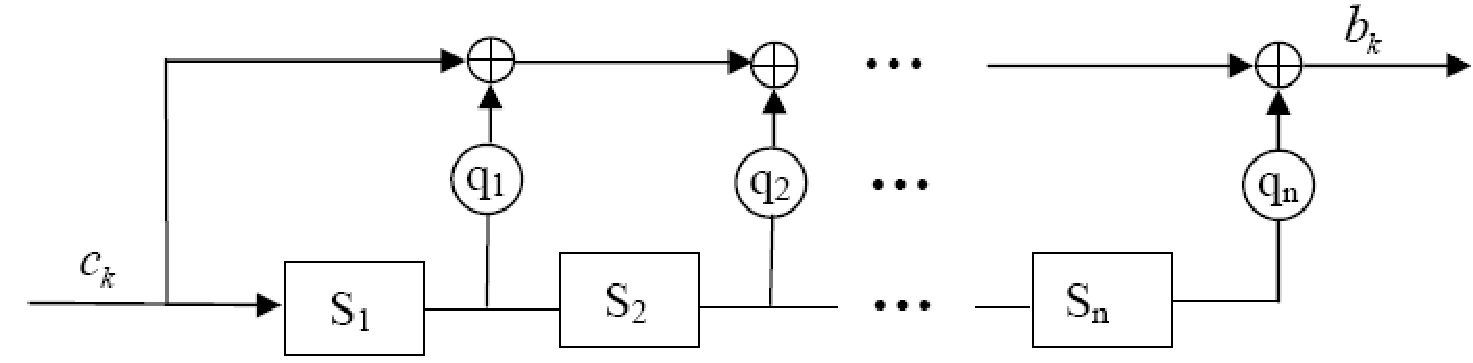}
\caption{Binary Decoder Structure of a FBC code generated by $g_{FBC}(x)$} \label{Bin_dec}
\label{fig10_0}
\end{figure}

Based on the above equation, we can derive the following binary
decoder structure in Fig. \ref{Bin_dec}, where the input is the codeword symbol $c_k$  and
the output is $b_k$.

Let  ${P_{{S_i}(k)}}(w)$ denote the probability of memory ${S_i}(k)
= w$  and ${\alpha _k}(m)$ denote the probability of state $m$ at
time $k$. Let $\left( {{m_1}, \cdots ,{m_n}} \right)$  be the
$n$-dimensional binary representation of $m$ and $\left( {{m'_1},
\cdots ,{m'_n}} \right)$ be the binary representation of $m'$. At
time $k$, with input $c_k$, the state transits from $\left( {{m'_1},
\cdots ,{m'_n}} \right)$ to $\left( {{m_1},{m_2}, \cdots ,{m_n}}
\right)$= $\left( {{c_k},{m'_1}, \cdots ,{m'_{n - 1}}} \right)$.
Then we have
\begin{eqnarray} \label{eq18}
{\alpha _k}(m) &=& \prod\limits_{i = 1}^n {{P_{{S_i}(k)}}({m_i})}  =
P({c_k} = {m_1})\prod\limits_{i = 2}^n {{P_{{S_i}(k)}}({m_i})} \\
\nonumber &=& P({c_k} = {m_1})\sum\limits_{{m'_n} = 0,1}^{}
{{P_{{S_n}(k -
1)}}({m'_n})\prod\limits_{i = 2}^n {{P_{{S_i}(k)}}({m_i})} } \\
\nonumber &=& \sum\limits_{{m'_n} = 0,1}^{} {\left( {\prod\limits_{j
= 1}^n {{P_{{S_j}(k - 1)}}({m'_j})} } \right)}
P({c_k} = {m_1}) \\
\nonumber &=& \sum\limits_{m'} {{\alpha _{k - 1}}(m'){\gamma
_k}(m',m)},
\end{eqnarray}
where ${\alpha _{k - 1}}(m') = \prod\limits_{j = 1}^n {{P_{{S_j}(k -
1)}}({m'_j})}$,  ${\gamma _k}(m',m)$=$P({c_k} = {m_1})$  and
${m'_j}$=$m_{j+1}$, for $j$=$1,2,\ldots,n-1$.

The APP of $b_k=w$ can then be calculated as
\begin{eqnarray} \label{eq19}
{p_{{b_k}}}(w) &=& p\left( {{b_k} = w|{\bf{y}}} \right) =
\sum\limits_{\left( {m',m} \right) \in U(b(k) = w)} {\prod\limits_{j
= 1}^n {{P_{{S_j}(k - 1)}}({m'_j})} P({c_k} = {m_1})} \\
\nonumber &=& \sum\limits_{{{m_1},{m'_1}, \cdots ,{m'_n}}, ~{{m_1}
\oplus \sum\limits_{j = 1}^n {{q_j}{m'_j} = w}}} {\prod\limits_{j =
1}^n {{P_{{S_j}(k - 1)}}({m'_j})} P({c_k} =
{m_1})} \\
\nonumber &=& \sum\limits_{{{m_1},{m'_1}, \cdots ,{m'_n}},~ {{m_1}
\oplus \sum\limits_{j = 1}^n {{q_j}{m'_j} = w} }} {\prod\limits_{j =
1}^n {P({c_{k - j}} = {m'_j})} P({c_k} =
{m_1})} \\
\nonumber &=& \sum\limits_{{m'_0},{m'_1}, \cdots ,{m'_n},~
{\sum\limits_{j = 0}^n {{q_j}{m'_j} = w}}}{\prod\limits_{j = 0}^n
{P({c_{k - j}} = {m'_j})} },
\end{eqnarray}
where ${m'_0} = {m_1}$ and ${q_0} = 1$.

Let $L({b_k})$ represent the LLR of $b_k$. From Eq. (\ref{eq19}) we
can easily derive
\begin{eqnarray} \label{eq20}
L({b_k}) = L\left( {\sum\limits_{j = 0}^n {{q_j}c_{k - j}^{}} }
\right).
\end{eqnarray}

Following the L-sum theory [7], the right-hand side of (\ref{eq20})
can be expanded as
\begin{eqnarray} \label{eq21}
L\left( {\sum\limits_{j = 0}^n {{q_j}{c_{k - j}}}
} \right) = \ln \frac{{1 + \prod\limits_{j = 0}^n {\tanh (L\left(
{{q_j}{c_{k - j}}} \right)/2)} }}{{1 - \prod\limits_{j = 0}^n {\tanh
(L\left( {{q_j}{c_{k - j}}} \right)/2)} }},
\end{eqnarray}
where $\tanh (x/2) = \frac{{{e^x} - 1}}{{{e^x} + 1}}$.

Then by using the following relationship between the LLR and soft
symbol estimate,
\begin{eqnarray} \label{eq22}
{\hat x_{b_k}} = \frac{{{e^{L({b_k})}} - 1}}{{{e^{L({b_k})}} +
1}}=\tanh\left(L(b(k)/2\right),
\end{eqnarray}
\begin{eqnarray} \label{eq23}
L({b_k}) = \ln \frac{{1 + {{\hat x}_{b_k}}}}{{1 - {{\hat
x}_{b_k}}}},
\end{eqnarray}
(\ref{eq20}) can be further written as
\begin{eqnarray} \label{eq24}
L({b_k}) = L\left( {\sum\limits_{j = 0}^n {{q_j}c_{k - j}^{}} }
\right) = \ln \frac{{1 + \prod\limits_{j = 0}^n {{{\hat
x}_{{q_j}{c_{k - j}}}}} }}{{1 - \prod\limits_{j = 0}^n {{{\hat
x}_{{q_j}{c_{k - j}}}}} }} = \ln \frac{{1 + \prod\limits_{j = 0}^n
{{{\left( {{{\hat x}_{c_{k - j}}}} \right)}^{{q_j}}}} }}{{1 -
\prod\limits_{j = 0}^n {{{\left( {{{\hat x}_{c_{k - j}}}}
\right)}^{{q_j}}}} }},
\end{eqnarray}
where ${\hat x_{{q_j}{c_{k - j}}}}$  denotes the soft symbol
estimate of symbol ${q_j}{c_{k - j}}$. Obviously ${\hat
x_{{q_j}{c_{k - j}}}} = 1$ when ${q_j} = 0$ and ${\hat x_{{q_j}{c_{k
- j}}}} = {\hat x_{c,k - j}}$ when ${q_j} = 1$. Thus ${\hat
x_{{q_j}{c_{k - j}}}} = {\left( {{{\hat x}_{c_{k - j}}}}
\right)^{{q_j}}}$.

By substituting (\ref{eq24}) into (\ref{eq22}), we get
\begin{eqnarray} \label{eq25}
{\hat x_{b_k}} = \prod\limits_{j = 0}^n {{{\left( {{{\hat x}_{c_{k -
j}}}} \right)}^{{q_j}}}}.
\end{eqnarray}

By taking the logarithm on both sides of (\ref{eq25}), we have
\begin{eqnarray} \label{eq26}
\ln {\hat x_{b_k}} = \sum\limits_{j = 0}^n {{q_j}\ln {{\hat x}_{c_{k
- j}}}}.
\end{eqnarray}

Therefore, the log-domain SISO forward decoding of the FBC code can
be simply implemented by its dual encoder, generated by the
generated polynomial ${q_{FBC}}(x) = 1/{g_{FBC}}(x) = {q_n}{x^n} +
\cdots {q_1}x + 1$.

This proved Theorem 1.

\subsection{Proof of Theorem 2}

\begin{figure}
\centering \subfigure[The binary decoder of FFC code generated by
$a(x)$, which is equivalent to an encoder generated by $1/a(x)$]{
\includegraphics[width=0.6\textwidth]{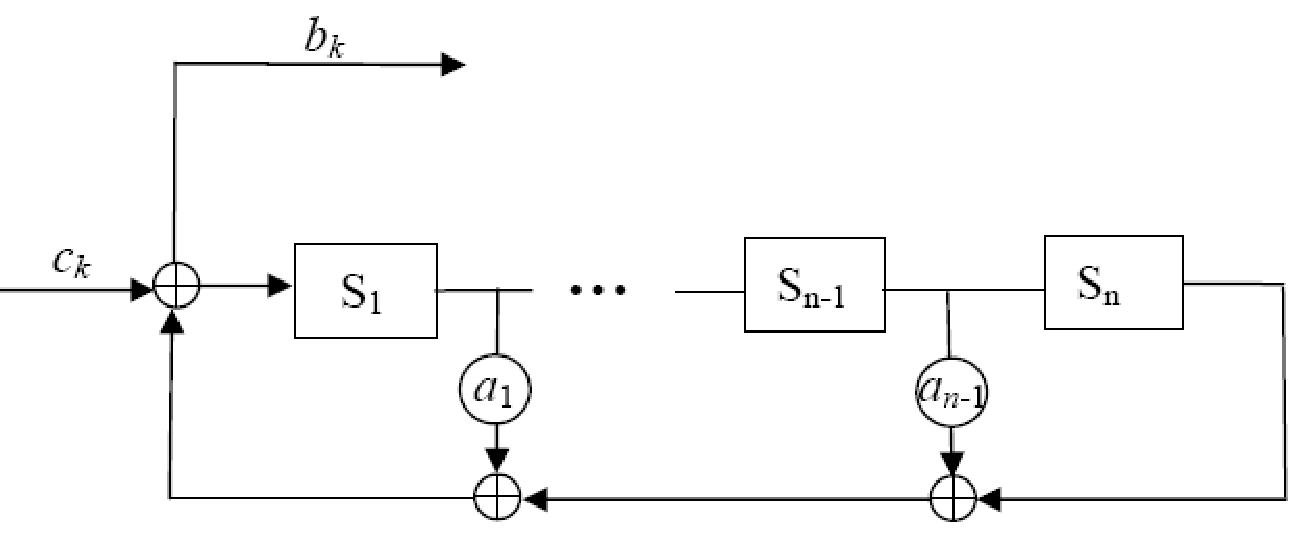}
\label{fig10_1} } \subfigure[The binary decoder of FFC code
generated by $\frac{a(x)z(x)}{z(x)}=\frac{x^{n+l}+1}{z(x)}$ , which
is equivalent to an encoder generated by
$\frac{z(x)}{a(x)z(x)}=\frac{z(x)}{x^{n+l}+1}$]{
\includegraphics[width=0.7\textwidth]{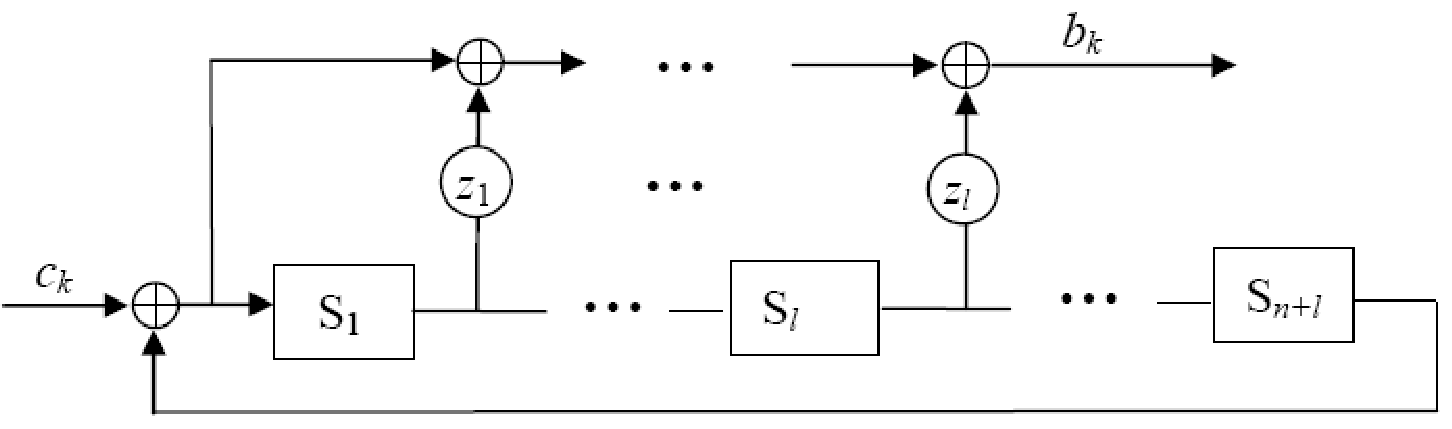}
\label{fig10_2}} \caption{The binary dual encoder of a FFC code}
\end{figure}

Let us first examine the forward binary decoding. Based on the code
generator polynomials, we can easily derive the binary decoder of
codes generated by  $a(x)$ and
$\frac{{a(x)z(x)}}{{z(x)}}$=$\frac{{{x^{n + l}} + 1}}{{z(x)}}$, as
shown in Fig. \ref{fig10_1} and \ref{fig10_2}, respectively. As can
be seen from these figures, the binary decoder of each of these two
codes is equivalent to the encoder generated by its respective
inverse polynomial.

Let $\left( {{m_1}, \cdots ,{m_n}} \right)$ and  $\left( {{m'_1},
\cdots ,{m'_n}} \right)$ be the $n$-dimensional binary
representation of $m$ and $m'$. Let $\left( {{u_1}, \cdots ,{u_{n +
l}}} \right)$ and $\left( {{u'_1}, \cdots ,{u'_{n + l}}} \right)$ be
the ($n+l$)-dimensional binary representation of $u$ and $u'$.
Assume that at time $k$, with input $c_k$, the state transits from
$\left( {{m'_1}, \cdots ,{m'_n}} \right)$ to $\left( {{m_1},m_2,
\cdots ,{m_n}} \right)$ in the binary decoder of Fig. \ref{fig10_1}
and transits from  $\left( {{{u'}_1}, \cdots ,{{u'}_{n + l}}}
\right)$ to $\left( {{u_1}, \cdots ,{u_{n + l}}} \right)$ in
\ref{fig10_2}. For a binary input sequence ${\bf{c}} = \left(
{{c_1},{c_2}, \cdots ,{c_K}} \right)$, it is well known that the
polynomials $\frac{1}{{a(x)}}$ and $\frac{{z(x)}}{{a(x)z(x)}}$
generate the same codeword. We thus have
\begin{eqnarray} \label{eq27}
{b_k} = \sum\limits_{i = 1}^{n - 1}{{a_i}m'_i + m'_n + {c_k}} =
\sum\limits_{j = 1}^{n + l - 1}{{z_j}u'_j + u'_{n + l} + {c_k}},
\end{eqnarray}
\begin{eqnarray} \label{eq28}
{m_j} = {m'_{j - 1}}, ~and ~{u_j} = {u'_{j - 1}}, j \ge 2.
\end{eqnarray}

Then by following similar calculation in Appendix A, we have
\begin{eqnarray} \label{eq29}
L({b_k}) = L\left( {\sum\limits_{i = 1}^{n - 1} {{a_i}m'_i + m'_n +
{c_k}} } \right),
\end{eqnarray}
\begin{eqnarray} \label{eq30}
L({b_k}) = L\left( {\sum\limits_{j = 1}^{n + l - 1} {{z_j}u'_j +
u'_{n + l} + {c_k}} } \right).
\end{eqnarray}

When the terms in the summation of the right-hand side in
(\ref{eq29}) and (\ref{eq30}) are statistically independent, we can
use the L-sum theory to further expand these two equations. However,
we can easily check that the terms $m'_i$, $i = 1, \cdots ,n$, in
(\ref{eq29}), are not independent. Now let us prove that $u'_i$, $i
= 1, \cdots ,n+l$ are statistically independent random variables.

When $0<k<n+l$, the state $u'_i$, $i = 1, \cdots ,n+l$,  at time
$k$, is given by
\begin{eqnarray} \label{eq31}
u'_i= 0, k<i~and~u'_i = {c_{k - i}}, k \ge i.
\end{eqnarray}

When $k>n+l$, the state $u'_i$, $i = 1, \cdots ,n+l$, at time $k$,
is given by
\begin{eqnarray} \label{eq32}
u'_i= \sum\limits_{p = 0}^{\lfloor{k/(n + l)}\rfloor} {{c_{k - pi}}},
\end{eqnarray}
where $\lfloor{x}\rfloor$ denotes the largest integer not greater
than $x$.

From (\ref{eq31}) and (\ref{eq32}), we can see that $u'_i$, $i = 1,
\cdots ,n+l$, are statistically independent random variables at any
time instant $k$.

\begin{figure}
\centering
\includegraphics[width=0.7\textwidth]{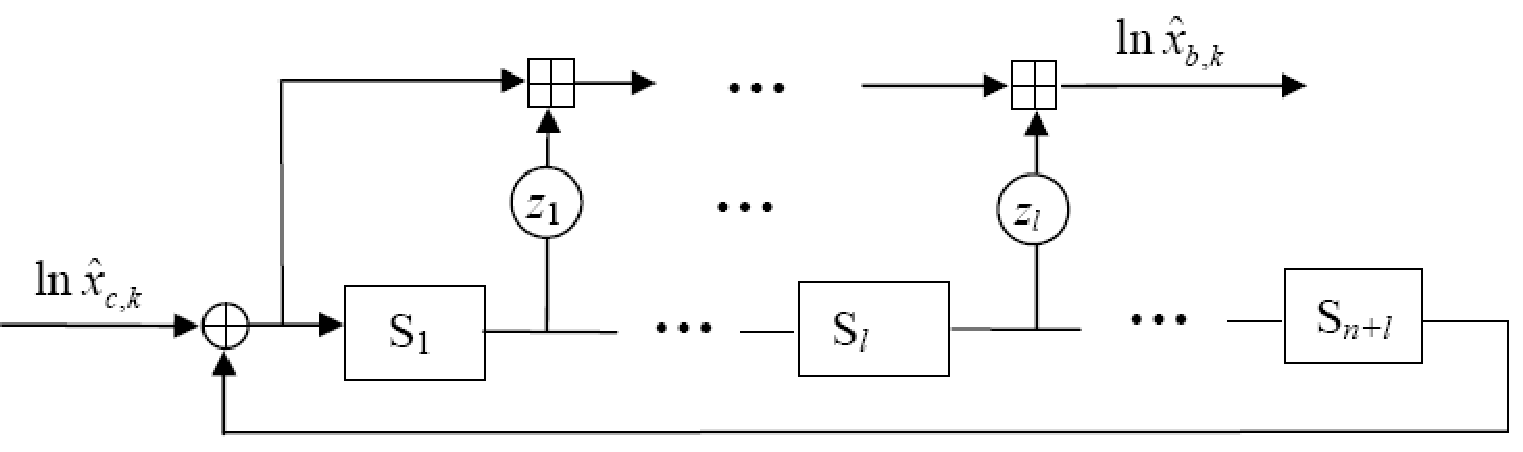}
\caption{The SISO decoder structure, implemented with the encoder
with the generator polynomial of
$q_{FFC}(x)=\frac{z(x)}{a(x)z(x)}=\frac{z(x)}{x^{n+l}+1}$}
\label{fig11}
\end{figure}

Since  $u'_i$, $i = 1, \cdots ,n+l$ are statistically independent
random variables, we can use the L-sum theory [7] to expand the
right-hand side of (\ref{eq31}). By following a similar calculation
as in Appendix A, we can obtain the following equation
\begin{eqnarray} \label{eq34}
\ln {\hat x_{b_k}} = \sum\limits_{j = 1}^{n + l - 1} {{z_j}{{\hat
x}_{u'_j}} + {{\hat x}_{u'_{n + l}}} + {{\hat x}_{c_k}}},
\end{eqnarray}
and,
\begin{eqnarray} \label{eq35}
{\hat x_{u_j}} = {\hat x_{u'_{j - 1}}}, j \ge 2,
\end{eqnarray}
where ${\hat x_{b_k}}$, ${\hat x_{u_j}}$, ${\hat x_{u'_j}}$ and
${\hat x_{c_k}}$ denotes the soft symbol estimate of symbol ${b_k}$,
${u_j}$, ${u'_j}$, and $c_k$, respectively. Based on (\ref{eq34})
and (\ref{eq35}), we can derive the SISO decoder structure, shown in
Fig. \ref{fig11}, implemented with the encoder with the generator
polynomial of
\begin{eqnarray}
{q_{FFC}}(x) = \frac{{z(x)}}{{a(x)z(x)}} = \frac{{z(x)}}{{{x^{n +
l}} + 1}}.
\end{eqnarray}

This proves Theorem 2.

\subsection{Proof of Theorem 3}

Assume that the encoder with the generator polynomial $g(x)$ in Fig.
\ref{fig6_1} transits from the state $\left( {{m'_1},{m'_2}, \cdots
,{m'_n}} \right)$ at time $k$-1 to the state $\left( {{m_1},{m_2},
\cdots ,{m_n}} \right)$ at time $k$ with input $b_k$, then we have
\begin{eqnarray} \label{eq36}
{m_1} = {b_k} + \sum\limits_{p = 1}^{n - 1} {{q_p}{m'_p}}  + {m'_n},
~{m_p} = {m'_{p - 1}},~ p \ge 2,
\end{eqnarray}
and the corresponding trellis output at time $k$ is given by
\begin{eqnarray} \label{eq37}
c(k) &=& {b_k} + \sum\limits_{p = 1}^{n - 1} {{q_p}{m'_p}}  +
{m'_n} + \sum\limits_{p = 1}^{m - 1} {{a_p}{m'_p}}  + {m'_m} \\
\nonumber  &=& {b_k} + \sum\limits_{p = 1}^{n - 1} {{q_p}{m'_p}} +
\sum\limits_{p = 1}^{m - 1} {{a_p}{m'_p}}  + {m'_n} + {m'_m}.
\end{eqnarray}

To prove Theorem 3, we now only need to prove that with input $b_k$
its \textit{reverse memory-labeling} encoder transits from the state
$\left( {{m_1},{m_2}, \cdots ,{m_n}} \right)$ at time $k$-1 to the
state $\left( {{m'_1},{m'_2}, \cdots ,{m'_n}} \right)$  at time $k$
and generate the same encoder output.

Now let us consider the \textit{reverse memory-labeling} encoder
with the generator polynomial $g(x)$ in Fig. \ref{fig6_2}. With the
state $\left( {{m_1},{m_2}, \cdots ,{m_n}} \right)$ at time $k$-1
and input $b_k$, the state at time $k$ of the \textit{reverse
memory-labeling} encoder is given by
\begin{eqnarray} \label{eq38}
{S_n}(k) = {b_k} + {m_1} + \sum\limits_{p = 1}^{n - 1} {{q_p}{m_{p +
1}}\mathop  = \limits^{(a)} } {b_k} + {b_k} + \sum\limits_{p = 1}^{n
- 1} {{q_p}{m'_p}}  + {m'_n} + \sum\limits_{p = 1}^{n - 1}
{{q_p}{m'_p} = } {m'_n},
\end{eqnarray}
\begin{eqnarray} \label{eq39}
{S_p}(k) = {S_{p + 1}}(k) = {m_{p + 1}} = {m'_p},
\end{eqnarray}
where in the step $(a)$ of (\ref{eq38}) we have used Eq.
(\ref{eq36}).

The output of \textit{reverse memory-labeling} encoder at time $k$
is given by
\begin{eqnarray} \label{eq40}
c(k) = {m_{n + 1}} + \sum\limits_{p = 1}^{m - 1} {{a_p}{m_{p + 1}}}
+ {m_1} = {m'_m} + \sum\limits_{p = 1}^{m - 1} {{a_p}{m'_p}}  +
{b_k} + \sum\limits_{p = 1}^{n - 1} {{q_p}{m'_p}}  + {m'_n},
\end{eqnarray}
where we have used Eq. (\ref{eq36}) in the last step of calculation.

From (\ref{eq38}-\ref{eq40}), we can see that with input $b_k$ the
reverse memory-labeling encoder transits from the state $\left(
{{m_1},{m_2}, \cdots ,{m_n}} \right)$ at time $k-1$ to the state
$\left( {{m'_1},{m'_2}, \cdots ,{m'_n}} \right)$ and generates the
same encoder output as the encoder with the generator polynomial
$g(x)$.

This proves Theorem 3.

\subsection{Proof of Theorem 5}

\begin{figure}
\centering
\includegraphics[width=0.7\textwidth]{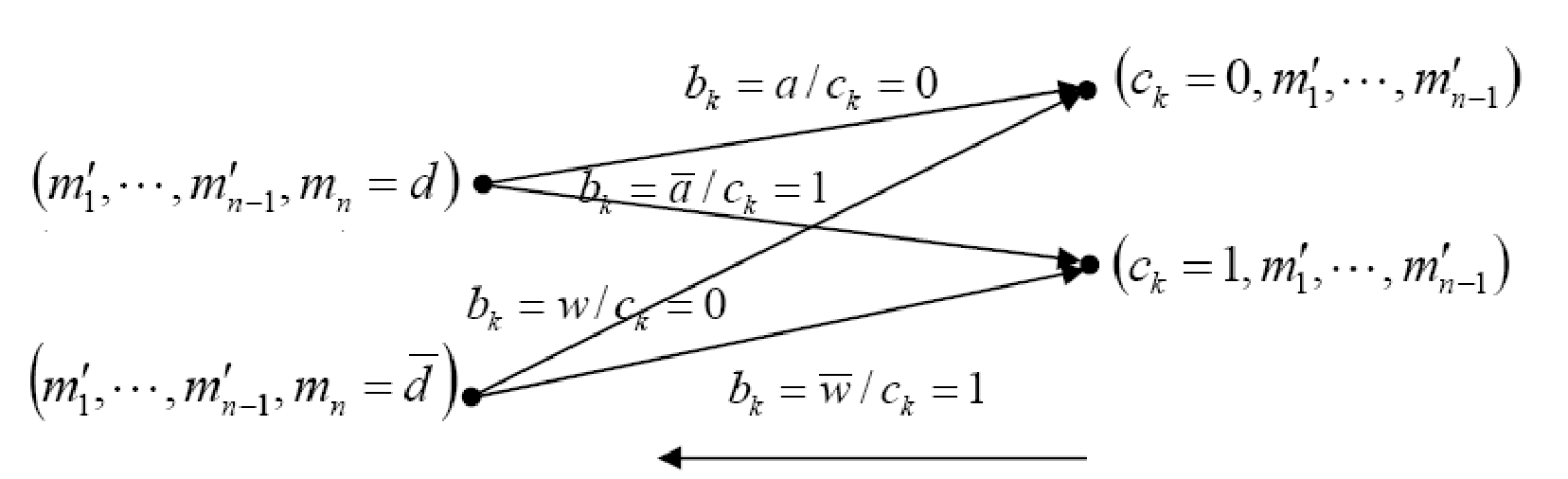}
\caption{The backward decoding trellis transition of a FBC code}
\label{fig12}
\end{figure}

To prove Theorem 5, let us first examine the backward decoding of a
FBC code. At the encoder of a FBC code in Fig. 5, with input $b_k$,
the state transits from  $\left( {{m'_1}, \cdots ,{m'_{n -
1}},{m'_n}} \right)$ at time $k-1$ to $\left( {{m_1},{m_2}, \cdots
,{m_n}} \right)$ = $\left( {{c_k},{m'_1}, \cdots ,{m'_{n - 1}}}
\right)$ at time $k$, where $c_k$ is the encoder output. The state
transition is shown in the Fig. \ref{fig12}, where $a$, $d$, $w$=0
or 1, $\bar a = 1 - a$, $\bar d = 1 - d$, and $\bar w = 1 - w$.

Then we apply the BCJR backward decoding as follows,

(1) At time $K$, we have ${\beta _K}(0) = 1$ and ${\beta _K}(m) = 0$
for $m \neq 0$;

(2) At time $K$-1, we have

${\beta _{K - 1}}(m) = {p_{{c_K}}}(0)$ for $m = 0,1$,~and~${\beta
_{K - 1}}(m) = 0$ for $m \neq 0,1$;

(3) At time $K$-2, we have

${\beta _{K - 2}}(m) = {p_{{c_{K - 1}}}}(0){p_{{c_K}}}(0)$, for $m =
0,1,2,3$, and ${\beta _{K - 2}}(m) = 0$, for $m \neq 0,1,2,3$;

$\vdots$

(4) At time $K-v$, $0 \le v \le n$, we have

${\beta _{K - v}}(m) = \prod\limits_{i = 0}^{v - 1} {{p_{{c_{K -
i}}}}(0)}$, for $m = 0,1 \cdots ,{2^v} - 1$, and ${\beta _{K -
v}}(m) = 0$, for $m \neq 0,1 \cdots ,{2^v} - 1$;

$\vdots$

At time $K-t$, $t>n$ , we have

${\beta _{K - t}}(m) = \prod\limits_{i = 0}^{n - 1} {{p_{{c_{K -
i}}}}(0)}$, for all $m$.

From the above equation, we can see that ${\beta _k}(m)$ is the same
for all states when $k \le K - n$. Therefore, the backward decoding
does not have any contribution in the probability calculation of the
BCJR decoding. This proves that the BCJR forward decoding is exactly
the same as the MAP decoding for the FBC codes.

\subsection{Proof of Lemma 1}\label{proof of lemma 1}
Let $S_i(k), i=1, 2, \cdots, n,$ and $S'_j(k), j=1, 2, \cdots, n+l,$
denote the memory of the $i$-th shift register of encoder C and the
$j$-th shift register of encoder $\bar{C}$, generated by $g_{GC}(x)$
and $q_{GC}(x)$. According to Fig. 7a, in encoder C, $S_1(k+1)$ is
given by
\begin{align}
S_1(k+1)=b_{k+1}+\sum_{i=1}^n q_i S_i(k),
\end{align}
equivalently, we have the following equation
\begin{align}\label{bk+1}
b_{k+1}=S_1(k+1)+\sum_{i=1}^n q_i S_i(k).
\end{align}

Let $q_0=1$ and $S_1(k+1)=S_0(k)$. Then (\ref{bk+1}) can be written
as
\begin{align}\label{encoder C}
b_{k+1}=\sum_{i=0}^n q_i S_i(k).
\end{align}

In encoder $\bar{C}$, the output $b_{k+1}$ can be written as
\begin{align}
b_{k+1}&=c_{k+1}+\sum_{j=1}^{n+l-1} h_jS'_j(k)\nonumber\\
&=S'_1(k+1)+\sum_{j=1}^{n+l-1} h_jS'_j(k)+S'_{n+l}(k),
\end{align}
where we have used the relationship of
$c_{k+1}=S'_1(k+1)+S'_{n+l}(k)$. Let $h_0=h_{n+l}=1$ and
$S'_1(k+1)=S'_0(k)$, and we get
\begin{align}
b_{k+1}&=\sum_{j=0}^{n+l} h_jS'_j(k).
\end{align}

Since $h(x)=q(x)z(x)=q(x)+z_1xq(x)+ \cdots +
z_{l-1}x^{l-1}q(x)+x^{l}q(x)$, we have
\begin{align}
b_{k+1}&=\sum_{i=0}^n q_i S'_i(k)+z_1 \sum_{i=0}^n q_i S'_{i+1}(k)+
\cdots +z_{l-1} \sum_{i=0}^n q_i S'_{i+l-1}(k)+\sum_{i=0}^n q_i
S'_{i+l}(k)\nonumber\\ \label{encoder C'} &=\sum_{i=0}^n q_i
\left\{S'_i(k)+z_1S'_{i+1}(k)+\cdots+z_{l-1}S'_{i+l-1}(k)+S'_{i+l}(k)\right\}.
\end{align}

Comparing (\ref{encoder C}) and (\ref{encoder C'}), we can represent
$S_i(k)$ by a linear combination of shift register memories of
encoder $\bar{C}$
\begin{align}
S_i(k)=S'_i(k) + z_1 S'_{i+1}(k) +z_2 S'_{i+2}(k)+ \cdots + z_{l-1}
S'_{i+l-1}(k)+S'_{i+l}(k).
\end{align}

Therefore, when tail bits of encoder $\bar{C}$ terminate it at the
all-zero state, these tail bits will also terminate encoder C at the
all-zero state.

This proves Lemma \ref{dual decoder and encoder states returen to
zero}.

\subsection{Proof of Lemma 2}\label{proof of lemma 2}
%

With tail bits, both encoder $\bar{C}$ and its backward encoder
begin with and end at the all-zero state, that is
\begin{align}
&\left(\overrightarrow{S'_1}(K+n+l), \overrightarrow{S'_2}(K+n+l), \cdots, \overrightarrow{S'_{n+l}}(K+n+l)\right)=\left(0, 0, \cdots, 0\right)\nonumber\\
&\left(\overleftarrow{S'_{1}}(K+n+l), \overleftarrow{S'_{2}}(K+n+l),
\cdots, \overleftarrow{S'_{n+l}}(K+n+l)\right)=\left(0, 0, \cdots,
0\right).
\end{align}

The state of encoder $\bar{C}$ and its backward encoder at time
$K+n+l-1$ can be calculated as
\begin{align}
&\left(\overrightarrow{S'_1}(K+n+l-1), \overrightarrow{S'_2}(K+n+l-1), \cdots, \overrightarrow{S'_{n+l-1}}(K+n+l-1), \overrightarrow{S'_{n+l}}(K+n+l-1)\right)\nonumber\\
&=\left(0, 0, \cdots, 0, c_{K+n+l}\right)\nonumber\\
&\left(\overleftarrow{S'_{1}}(K+n+l-1), \overleftarrow{S'_{2}}(K+n+l-1), \cdots, \overleftarrow{S'_{n+l-1}}(K+n+l-1), \overleftarrow{S'_{n+l}}(K+n+l-1)\right)\nonumber\\
&=\left(0, 0, \cdots, 0, c_{K+n+l}\right).
\end{align}
This means that the encoder $\bar{C}$ and its backward encoder will
arrive at the same state at time $K+n+l-1$. Let ${\bf
u'}=\left(u'_1, u'_2, \cdots, u'_{n+l}\right)$ denote the state of
encoder $\bar{C}$ at time $k-1$. Then its next state at time $k$ is
given by ${\bf u}=\left(u_1, u_2, \cdots,
u_{n+l}\right)=\left(c_k+u'_{n+l}, u'_1, \cdots, u'_{n+l-1}\right)$.
To prove that the encoder $\bar{C}$ and its backward encoder arrive
at the same state any any time $k$, we only need to prove that the
backward encoder will transit from state ${\bf u}$ at time $k$ to
state ${\bf u'}$ at time $k-1$. This can be proved in a similar way
as the proof of Theorem 3 and we omit it here.

%
%
%
%

This proves Lemma \ref{same state transitions}.

\subsection{Proof of Theorem 6}\label{proof of theorem 1}
We consider a GC code generated by $g_{GC}(x)=\frac{a(x)}{q(x)}$.
Its dual encoder for decoding is described by
$q_{GC}(x)=1+\frac{h_1x+ \cdots + h_{n+l-1}x^{n+l-1}}{1+x^{n+l}}$.
We assume that the state of the dual encoder $\bar{C}$ transits from
$\left(u'_1, u'_2, \cdots, u'_{n+l}\right)$  at time $k-1$ to
$\left(u_1, u_2, \cdots, u_{n+l}\right)$ at time $k$ with input
$c_k$. According to $q_{GC}(x)$, the output of the dual encoder
$\bar{C}$ at time $k$ can be written as
\begin{align}
b_k=c_k+\sum_{j=1}^{n+l-1}h_ju'_j.
\end{align}
%
%

For the bidirectional BCJR MAP algorithm, the probability that
$b_k=w$ is given by
\begin{align}\label{pbk}
&P_{b_k}(\omega)=P\left\{b_k=\omega|\vec{y}\right\}=\sum_{(u',u)=U(b_k=\omega)}\alpha_{k-1}\left(u'\right)\gamma_k\left(u', u\right)\beta_k(u)\nonumber\\
&=\sum_{(u',u)=U(b_k=\omega)}\prod_{j=1}^{n+l-1} P_{\overrightarrow{S'}_{j}(k-1)}(u'_j)P\left( c_k\right)\prod_{i=2}^{n+l}P_{\overleftarrow{S'}_{i}(k)}(u_i)\nonumber\\
&=\sum_{(u',u)=U(b_k=\omega)}\prod_{j=1}^{n+l-1} P_{\overrightarrow{S'}_{j}(k-1)}(u'_j)\prod_{j=1}^{n+l-1}P_{\overleftarrow{S'}_{j}(k-1)}(u'_j)P\left(c_k \right)\nonumber\\
&=\sum_{c_k, u'_1, \cdots, u'_{n+l},
\sum_{j=1}^{n+l-1}h_ju'_j+c_k=\omega}\prod_{j=1}^{n+l-1}
P_{S'_{j}(k-1)}(u'_j)P\left(c_k \right),
\end{align}
where $P_{S'_{j}(k-1)}(u'_j)=P_{\overrightarrow{S'}_{j}(k-1)}(u'_j)
P_{\overleftarrow{S'}_{j}(k-1)}(u'_j)$. Let $L\left(b_k\right)$
denote the LLR of $b_k$. From (\ref{pbk}), we can get
\begin{align}
L\left(b_k\right)=L\left(\sum_{j=1}^{n+l-1} h_j
S'_j(k-1)+c_k\right),
\end{align}
where $L\left\{S'_j(k-1)\right\}=\ln
\frac{P_{S'_{j}(k-1)}(0)}{P_{S'_{j}(k-1)}(1)}
=\overrightarrow{L}_{S'_j}(k-1)+\overleftarrow{L}_{S'_j}(k-1)$. Let
$\hat{V}_j(k-1)$ denote the SSE of $S'_j(k-1)$, and we can get
\begin{align}\label{xbk}
\ln \hat{x}_{b_k}= \ln \hat{x}_{c_k}+\sum_{i=1}^{n+l-1} h_i \ln
\hat{V}_{i}(k-1).
\end{align}

Comparing the shift register combined outputs of the dual encoder
(\ref{shift register combined output of the dual encoder}) and the
outputs of the bidirectional BCJR MAP algorithm (\ref{xbk}), we can
see that they are exactly of the same.

\bibliographystyle{IEEE}

\begin{thebibliography}{25}

\bibitem{1}
P. Elias, "Coding for noisy channels," IRE Conv. Rec., pp. 4:37-47,
1955.

\bibitem{2}
A. Viterbi, "Error bounds for convolutional codes and an
asymptotically optimum decoding algorithm," IEEE Trans. Inform.
Theory, vol. 15, pp. 260-269, Apr. 1967.

\bibitem{3}
G. D. Forney, "The Viterbi algorithm",  Proc. of IEEE,  vol. 61,
pp.268 - 278 , 1973.

\bibitem{4}
G. D. Forney, "Convolutional codes II: Maximum-likelihood decoding,"
Inform. Control, vol. 25,  pp.222 - 250 , 1974.

\bibitem{5}
J. Hagenauer and P. Hoeher, "A Viterbi algorithm with soft-decision
outputs and its applications," Proc. IEEE GLOBECOM, pp. 47.11-47.17,
Dallas, TX, Nov 1989.

\bibitem{6}
L. Bahl, J. Cocke, F. Jelinek, and J. Raviv, "Optimal decoding of
linear codes for minimizing symbol error rate," IEEE Transactions on
Information Theory,  vol. 20, no. 2, pp. 284 - 287, Mar 1974.

\bibitem{7}
J. Hagenauer, E. Offer, and L. Papke, "Iterative decoding of binary
block and convolutional codes," IEEE Trans. Inf. Theory, vol. 42,
no. 2, pp. 429-445, Mar. 1996.

\bibitem{8}

Y.Li, B. Vucetic, and Y. Sato, "Optimum soft-output detection for
channels with intersymbol interference," IEEE Trans. Inf. Theory,
vol. 41, no. 3 March 1995, pp. 704 - 713.

\bibitem{9}

Y. Li, M. S. Rahman, and B. Vucetic, ”SISO MAP decoding of rate-1 recursive convolutional codes: a revisit,” Prof. ISIT
2012, MIT, Boston June 2012.

\end{thebibliography}

%

\end{document}